\documentclass[aps,pra,twocolumn,superscriptaddress]{revtex4}
\usepackage{amsfonts,epsfig}
\usepackage{amsmath}
\usepackage{amssymb}
\usepackage{amsfonts}
\usepackage{amsthm}
\usepackage{mathrsfs}
\usepackage{color,verbatim,graphics}
\usepackage{url}
\usepackage{multirow}

\newcommand{\ket}[1]{|#1\rangle}
\newcommand{\bra}[1]{\langle#1|}
\newcommand{\ketbra}[2]{|#1\rangle\langle#2|}
\newcommand{\bracket}[3]{\langle#1|#2|#3\rangle}
\newcommand{\braket}[2]{\langle#1|#2\rangle}
\newcommand{\tr}[1]{\textrm{Tr}\left(#1\right)}
\newcommand{\ave}[1]{\langle#1\rangle}

\renewcommand{\SS}{\mathcal{S}}

\definecolor{nblue}{rgb}{0.2,0.2,0.7}
\def\bea{\begin{eqnarray}}
\def\eea{\end{eqnarray}}
\def\bma{\begin{mathletters}}
\def\ema{\end{mathletters}}

\def\Q{{\cal Q}}
\def\0{\overline{0}}

\def\q0{\underline{0}}

\def\H{{\cal H}}

\def\C{{\mathbb C}}
\def\id{{\mathbb I}}

\def\H{{\cal H}}

\def\B{{\cal B}}

\def\tr{\mbox{tr}}
\def\proj#1{\ket{#1}\!\bra{#1}}
\def\Re{\mbox{Re}}
\def\cc{{d}} 

\newcommand{\be}{\begin{equation}}
\newcommand{\ee}{\end{equation}}
\newcommand{\ba}{\begin{eqnarray}}
\newcommand{\ea}{\end{eqnarray}}
\newcommand{\ban}{\begin{eqnarray*}}
\newcommand{\ean}{\end{eqnarray*}}

\newtheorem*{theorem*}{Theorem}

\renewcommand{\SS}{\mathcal{S}}

\begin{document}

\title{Physical characterization of quantum devices from nonlocal correlations}

\author{Jean-Daniel Bancal}
\affiliation{Centre for Quantum Technologies, National University of Singapore, 3 Science drive 2, Singapore 117543}
\author{Miguel Navascu\'es}
\affiliation{School of Physics, University of Bristol, Tyndall Avenue, Bristol BS8 1TL (U.K.)}
\author{Valerio Scarani}
\affiliation{Centre for Quantum Technologies, National University of Singapore, 3 Science drive 2, Singapore 117543}
\affiliation{Department of Physics, National University of Singapore, 2 Science drive 3, Singapore 117542}
\author{Tam\'as V\'ertesi}
\affiliation{Institute for Nuclear Research, Hungarian Academy of Sciences, H-4001 Debrecen, P.O. Box 51, Hungary}
\author{Tzyh Haur Yang}
\affiliation{Centre for Quantum Technologies, National University of Singapore, 3 Science drive 2, Singapore 117543}

\begin{abstract}
In the device-independent approach to quantum information
theory, quantum systems are regarded as black boxes which, given
an input (the measurement setting), return an output (the
measurement result). These boxes are then treated regardless of
their actual internal working. In this paper, we develop SWAP, a
theoretical concept which, in combination with the tool of
semi-definite methods for the characterization of quantum
correlations, allows us to estimate physical properties of
the black boxes from the observed measurement statistics. We find that the SWAP tool provides bounds orders of magnitude better than
previously-known techniques (e.g.: for a CHSH violation larger than 2.57, SWAP predicts a singlet fidelity greater than $70\%$). This method also allows us to deal with
hitherto intractable cases such as robust device-independent self-testing of
non-maximally entangled two-qutrit states in the CGLMP scenario
(for which Jordan's Lemma does not apply) and the device-independent certification of entangled
measurements. We further apply the SWAP method to relate
nonlocal correlations to work extraction and quantum dimensionality,
hence demonstrating that this tool can be used to study a wide variety of
properties relying on the sole knowledge of accessible
statistics.

\end{abstract}

\maketitle


\section{Introduction}
\noindent In the last few years, several tasks for which quantum
physics offers an advantage over classical methods have been found
possible without assuming knowledge of the inner working of the
devices involved, i.e. in a device-independent manner. This is the
case of key distribution~\cite{qkd1,qkd2,qkd3,qkd4,qkd5} and
randomness certification~\cite{random1,random2} for instance (see
also~\cite{review,slovakia} and references therein for more
examples).

In a device-independent experiment, also known as a Bell-type
experiment, conclusions are not drawn from the knowledge that specific
resources are used by the devices -- the content of the devices might
not be known a priori --, but only from the observed relation between
their inputs and outputs. Therefore, such devices are referred to
as black boxes. Any actual physical realization of black boxes
providing an advantage over classical devices must however use
some quantum resource. For instance, the violation of a bipartite
Bell inequality requires two black boxes to share an entangled
quantum state. Some knowledge about the underlying quantum state
can thus be inferred from the observed correlations only.

\begin{figure}
{\centering\includegraphics[width=0.45\textwidth]{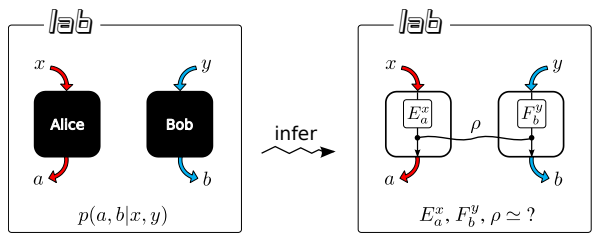}}
\caption{We are interested in certifying the physical properties
of black boxes. This amounts to question the nature of the
underlying quantum state $\rho$ and of the measurement operators
$E^x_a$, $F^y_b$ which are necessary for some boxes to produce the
observed statistics $p(a,b|x,y)$.} \label{fig:setup}
\end{figure}

Several works have shown that, beyond the mere presence of entanglement, other
quantum features such as genuine entanglement~\cite{bancal11},
the dimension of the underlying Hilbert space~\cite{brunner08}, or the overlap
between measurements~\cite{tomamichel13} can also be witnessed
from correlations only. Moreover, quantitative statements have
also been obtained, for instance on the amount of
entanglement~\cite{bardyn,moroder}. But even quantifying these
aspects never fully characterizes the underlying quantum
realization:
several states have the same negativity or Hilbert space
dimension, and several sets of measurements can have the same
overlap. In fact, all of these estimations could be deduced if one
were able to identify which states and measurements are performed
by the black boxes in the first place. This is the subject of this
paper (see Figure~\ref{fig:setup}).

The question of identifying quantum states and measurements
from correlations only was originally addressed in two inequivalent ways. The
first criterion is due to Refs.~\cite{Tsirelson93,sw87,pr92},
where it is shown that if the Clauser-Horne-Shimony-Holt
(CHSH) inequality \cite{chsh} is violated maximally (given by the
famous value of $2\sqrt 2$ \cite{tsirelson}), then the state being
measured is equivalent to a maximally entangled state and the
measurements are anti-commuting on both Alice's and Bob's
part. Another criterion that certifies the same state and
measurements is due to Mayers and Yao \cite{yao} (see also
\cite{werner}).

These pioneering works showed that there exist lists of
statistical data in the device-independent framework which allow
one to identify the quantum state and measurement operators
involved in the actual experiment. Moreover, the process can tolerate a
small amount of external noise. This notion of determining
approximately the state and measurement operators involved via
non-locality detection is known as \emph{device-independent self-testing} (or simply \emph{self-testing} for short), and it has
inspired a number of works on the subject
\cite{st1,mckague,miller,reichardt,yang}.

However, the weak noise tolerance exhibited by the self-testing
protocols proposed so far makes them inapplicable in realistic
experimental situations. Moreover, all of them are specific to a
given Bell inequality \cite{st1,mckague,reichardt}, or family of
Bell inequalities \cite{yang}, and thus cannot be applied to
arbitrary states or measurements.

In this paper, building on the method sketched in
Ref.~\cite{yang_prl}, we present the SWAP method, a unified tool
that allows one to guess the state, measurements or other general
properties of the quantum systems inside the boxes from the
experimentally accessible correlations. The process is
algorithmic, i.e., it is completely automated, and makes use of
the Navascu\'es-Pironio-Ac\'in (NPA) hierarchy for the
characterization of quantum correlations \cite{navascues}. To
illustrate the power of this method, we apply it to estimate
different physical properties in standard Bell scenarios.

The structure of this paper is as follows. First, we recall the
formal definition of self-testing. Next, in Section \ref{method},
we explain the main idea behind our method: an effective
description, in the language of moment matrices, of the process of
exchanging the content of black boxes with a trusted
finite-dimensional system. This idea, together with the NPA
hierarchy \cite{navascues,navascueslong}, allows us to estimate
quantum monotones via convex optimization \cite{convex}, in the
same way that we would calculate such monotones had we known the
exact mathematical representation of the state and measurement
operators involved in the non-locality experiment. We also
indicate in this section how our results can be easily generalized
to the case where black boxes are not independently and
identically distributed (i.i.d.). In Section \ref{more_robust}, we
demonstrate that the method improves the noise tolerance of
previous self-testing protocols by several orders of magnitude.
Next, we show how the method can be applied to Bell-type scenarios
for which, due to their complexity, no robust device-independent
protocols (no matter how fragile to noise) could be derived
before. These are: self-testing of states with measurements having
more than two possible outcomes (Section \ref{CGLMP_app}); the
certification of entangled measurements (Section
\ref{entangling}); and the estimation of the amount of work
extractable from non-local quantum systems (Section \ref{work}).

\section{Self-testing}

Let Alice and Bob be two distant observers conducting measurements
on a shared quantum system. If Alice (Bob) makes no assumption on
the inner mechanisms of their experimental devices, she (he) has
to regard each of her (his) device as a \emph{black box} where she
(he) inputs a symbol $x$ ($y$) -- the measurement setting -- and
obtains an output $a$ ($b$) -- the measurement outcome. In these
conditions, Alice and Bob can make a frequency analysis to
estimate $p(a,b|x,y)$, the probability that they observe results
$a,b$ when they input $x,y$. The question of self-testing is
whether the user can deduce, from the knowledge of these
conditional probabilities only, the state and measurements used
inside their boxes (see Figure~\ref{fig:setup}).

According to quantum theory, there exist a state $\rho$ and measurement operators $\{E^x_a,F^y_b\}$ such that $p(a,b|x,y)=\tr(\rho\, E^x_a\otimes F^y_b)$. Assuming that all \emph{physical} implementations of a Positive Operator Valued Measure (POVM) involve a projective measurement over a dilated space, the operators $\{E^x_a,F^y_b\}$ can be taken as projectors. The underlying quantum state $\rho$ to which both parties have access in their labs can, in principle, be mixed. For simplicity, for all subsequent discussions, we will take it to be pure, i.e., $\rho=\proj{\Psi}$. However, as the reader can appreciate, all results presented in this paper hold independently of this assumption.

Now, suppose that, after many runs of an experiment, the parties
observe a probability distribution that is compatible with some
\emph{specific} state and measurements $(\ket{\overline{\psi}},
\{\overline{E}^x_a, \overline{F}^y_b\})$ defined in
$\C^d\otimes\C^d$: \be\label{eq:guessedRealization}
p(a,b|x,y)=\bra{\overline{\psi}}\overline{E}^x_a\otimes
\overline{F}^y_b\ket{\overline{\psi}}. \ee In general, this
observation does not imply that the actual state $\ket{\psi}$ and
measurements $\{E^x_a,F^y_b\}$ used by the black boxes are these
ones: since correlation
admit in general several possible quantum realizations, the actual
states and measurements could be different. Remarkably, however,
for some $p(a,b|x,y)$ it turns out that the guessed state and
measurements $(\ket{\overline{\psi}}, \{\overline{E}^x_a,
\overline{F}^y_b\})$ are the \textit{only} possible ones, up to
local unitaries and the addition of irrelevant degrees of freedom.
In this case, the observation of $p(a,b|x,y)$ certifies that the
states and measurements used by the boxes are the guessed ones.

Technically, this is captured by the following
\textit{self-testing statement}: the correlations $p(a,b|x,y)$
allow for self-testing if for every quantum realization
$(\ket{\psi}, \{E^x_a, F^y_b\})$ compatible with $p(a,b|x,y)$
there exists a local isometry
$\mathbf{\Phi}=\mathbf{\Phi}_A\otimes\mathbf{\Phi}_B$ such that
\ba
\mathbf{\Phi}\ket{\psi}&=&\ket{\overline{\psi}}_{A'B'}\otimes\ket{\textrm{junk}}_{AB}\label{ststate}\\
\mathbf{\Phi}(E^x_a\otimes
F^y_b\ket{\psi})&=&(\overline{E}^x_a\otimes\overline{F}^y_b\ket{\overline{\psi}})_{A'B'}\otimes\ket{\textrm{junk}}_{AB}\label{stmeas}
\ea where $A'$ and $B'$ are two ancillary qudits. The isometry must be
seen as a \textit{virtual protocol}: it does not need to be
implemented in the lab as part of the procedure of self-testing,
all that must be done in the lab is to query the boxes and derive
$p(a,b|x,y)$.

Intuitively, the isometry consists of \textit{local swaps} that put the relevant degrees of freedom from the boxes into the ancillas, which have the right dimension for a comparison with the desired state and measurement to be possible. Indeed, the intuition of the swap was already mentioned in the work of Mayers and Yao \cite{yao}. However, it was not exploited to derive a general method as we are going to do here.

\section{The method}
\label{method}

We first present the SWAP method with a specific example and then show
how a general construction is possible for any self-testing
scenario. Finally we show how to remove the assumptions of
infinitely many runs and the assumption that the behavior of the
devices is the same in each run and uncorrelated among the runs
(the latter being usually called \textit{i.i.d. assumption} by
analogy with sampling of probability distributions).

\subsection{Example: self-testing of qubit states}
\label{sec:exqubit}

\subsubsection{Construction of the swap operator}
\label{sec:swapqubit}

Since the swaps are local operations, it is enough to show how to construct one for Alice (the same construction applies to the other parties).
In case the state we wish to self-test is a qubit state, the local
ancilla $A'$ is a qubit. To construct the swap, we start by
temporarily assuming that the state inside the box is indeed a
qubit, and then relax this assumption.

In case the state inside Alice's box is a qubit, the swap operator
between her system and ancilla can be written as the concatenation
of three CNOT gates with alternate controls: \ba {\cal
S}_{AA'}&=&UVU\,,\label{uvu} \label{eq:SwapUVU} \ea where \be
\begin{array}{lcl}
U&=&\mathbb{I}\otimes\ket{0}\bra{0}+\sigma_x\otimes \ket{1}\bra{1}\\
V&=&\ket{0}\bra{0}\otimes \mathbb{I}+\ket{1}\bra{1}\otimes \sigma_x
\end{array}
\ee
are the CNOT gates with the ancilla, respectively the system, as control.

For simplicity, in this introduction we assume further that the
operators $\sigma_z$ and $\sigma_x$ are part of the set of
operators we want to self-test (see Section \ref{partialqubit} for an
example when this is not the case). For instance, \ba
\sigma_z&= & \overline{A}_0\,\equiv\,\overline{E}^0_{a=+1}-\overline{E}^0_{a=-1}\\
\sigma_x&= & \overline{A}_1\,\equiv\,\overline{E}^1_{a=+1}-\overline{E}^1_{a=-1}\,.
\ea
The trick is now to rewrite $U$ and $V$ with $\overline{A}_0$ and $\overline{A}_1$ on the side of the system, then to replace them by the real $A_0$ and $A_1$ operators defined analogously from the actual projectors $E^x_a$. This gives
\be
\begin{array}{lcl}
U&=&\mathbb{I}\otimes\ket{0}\bra{0}+A_1\otimes \ket{1}\bra{1}\\
V&=&\frac{\mathbb{I}+A_0}{2}\otimes \mathbb{I}+\frac{\mathbb{I}-A_0}{2}\otimes \sigma_x\,.
\end{array}
\label{uvqubit}\ee At this point, we have dropped the requirement
that the system is a qubit. Nevertheless, these two operators are
still unitary, because, by assumption, $A_x^2=\mathbb{I}$.
Finally, the guess for the swap operator is just \eqref{uvu} with
the new expressions~\eqref{uvqubit} for $U$ and $V$.

Using the same construction for Bob's side, the final
\textit{guess for the swap operator} is \ba\label{defS} {\cal
S}&=&{\cal S}_{AA'}\otimes {\cal S}_{BB'}\,. \ea

A few remarks are in order here. First, note that ${\cal
S}_{AA'}=VUV$ would be an equally valid choice instead
of~\eqref{eq:SwapUVU}. However, in several situations it can
be preferable to use~\eqref{eq:SwapUVU}. For instance, if the
ancilla is initialized in the state $\ket{0}$, then the action of
the first $U$ is equivalent to the identity. $\mathcal{S}_{AA'}$
then reduces to $UV$, which leads to simpler numerical
computations than $VUV$. We refer to $UV$ as a partial swap.

Second, note that the swap defined here explicitly depends on the
measurement operators used by the individual parties. This
guarantees that the result of applying this swap to the measured
state is identical for all unitarily equivalent quantum
representations.

\subsubsection{State certification with the SDP hierarchy}

Once the swap operator is defined, the next step of the method
describes how to use the knowledge of the correlations
$p(a,b|x,y)$.

For this, consider the self-testing of the state \eqref{ststate};
the self-testing of measurements \eqref{stmeas} follows a similar
path and will be sufficiently illustrated by our example in
Section \ref{measqubit}. A possible figure of merit is the
\textit{fidelity} \ba
F&=&\bracket{\overline{\psi}}{\rho_\text{swap}}{\overline{\psi}}
\label{fide_CHSH}\ea where $\rho_\text{swap}$ is the state of the
ancillas after the swap. The relation between this figure of merit
and the trace-distance between the desired and actual states is
discussed in Section~\ref{CHSH_snglt}.

Explicitly, if the initial dummy states of the ancillas are
denoted by $\rho_{A'}$ and $\rho_{B'}$, we have
\begin{equation}\label{eq:swappedstate}
\rho_\text{swap}=\tr_{AB}\left[{\cal S}\,
(\ket{\psi}\bra{\psi})_{AB}\otimes\rho_{A'}\otimes\rho_{B'}\,{\cal
S}^\dagger\right]\,,
\end{equation}
where the SWAP operator ${\cal S}$ is defined by
Eq.~(\ref{defS}). Having expressed ${\cal S}$ in terms of four of
the desired operators $A_0$, $A_1$, $B_0$ and $B_1$, we find $\rho_\text{swap}$ to be a $4\times 4$ matrix whose entries are linear combinations of expectation
values like $\bracket{\psi}{A_x\otimes\mathbb{I}}{\psi}$,
$\bracket{\psi}{A_0 A_1\otimes\mathbb{I}}{\psi}$,
$\bracket{\psi}{A_x\otimes B_y}{\psi}$, etc. Consequently, our
estimate $f$ on the singlet fidelity $F$ attainable via isometries
(\ref{fide_CHSH}) is a linear combination of operator averages of
the form $c_t=\bra{\psi}t(A_x,B_y)\ket{\psi}$, where $t$ is a
product of the operators $A_x,B_y$.

Self-testing is perfect, i.e. \eqref{ststate} holds exactly,
if we find $f=1$; but the method automatically deals with
imperfect cases, that is, it is robust by construction. So we are
finally left to estimate \textit{the minimal possible value of $f$
compatible with $p(a,b|x,y)$ and with quantum physics}. In other
words, we must solve the problem

\begin{align}
    \underset{d}{\min} \;\; &f(\cc)=\bracket{\overline{\psi}}{\rho_{\mathrm{swap}}(\cc)}{\overline{\psi}} \nonumber\\
    \textrm{s.t. }&\cc\in {\cal Q} \nonumber\\
                & \frac{1}{4}\left\{1+(-1)^a\cc_{A_x}+(-1)^b\cc_{B_y}\right.\nonumber\\
                &\ \ \left. +\ (-1)^{a+b}\cc_{A_xB_y}\right\}=p(a,b|x,y).
                \label{fide_Q}
\end{align}

\noindent where the maximization $\cc\in {\cal Q}$ is performed over
the set $\Q$ of all vectors $\cc=(\cc_\id,\cc_{A_0},...)$ that admit a
quantum representation, i.e., for which there exists a state
$\ket{\psi}$ and dichotomic operators $A_x,B_y$ such that
$\cc_t=\bra{\psi}t(A_x,B_y)\ket{\psi}$ for all products $t$.
Note that in the optimization problem above we assumed
for simplicity a probability distribution $p(a,b|x,y)$ with
dichotomic settings, $a,b\in \{0,1\}$.

Optimizations over the set of quantum momenta are computationally
hard \cite{ito}, and, in some scenarios, conjectured to be
undecidable \cite{tobias}. That is why we propose to relax the
above problem to an optimization over the sets $\Q^S$, defined in
\cite{navascues} as outer approximations of the quantum set. Let
$S$ be a set of products of the operators $A_x,B_y$, and let $\cc$
be a complex vector whose entries are labeled by products of the
form $s^\dagger t$, with $s,t\in S$ (objects such as $\cc$ will be
called \emph{moment vectors}). We define $\Gamma^S(\cc)$ as a matrix
whose rows and columns are numbered by products belonging to $S$
and such that $\Gamma_{s,t}=\cc_{s^\dagger t}$. It can be shown that
$\cc\in \Q$ implies that $\Gamma^S(\cc)$ is positive semidefinite. The
relaxation we propose to attack problem (\ref{fide_Q}) is thus:

\begin{align}
    f^S=\underset{d}{\min} \;\; &f(\cc) \nonumber\\
    \textrm{s.t. }&\Gamma^S(\cc)\geq 0 \nonumber\\
                & \frac{1}{4}\left\{1+(-1)^a\cc_{A_x}+(-1)^b\cc_{B_y}+\right.\nonumber\\
                &\ \ \left. +\ (-1)^{a+b}\cc_{A_xB_y}\right\}=p(a,b|x,y).
                \label{fide_Qn}
\end{align}

\noindent This last problem can be formulated as a semidefinite
program (SDP)~\cite{convex}, a type of convex optimization for
which there exist efficient numerical solvers to find global
minima and which also return sound error bounds on the optimal
guess. A note on notation: whenever $S$ corresponds to the set of
all products of $n$ or less measurement operators, we will denote
the corresponding set $\Q^S$ as $\Q^n$. It can be verified that,
if certain conjectures in von Neumann algebras hold \cite{ozawa},
the sequence of sets $(\Q^n)_n$ converges to $\Q$
\cite{navascueslong}.

We stress that the bound on $F$ obtained through this method 
is a bound not only on the fidelity of the swapped state with respect to
the target state $\ket{\overline{\psi}}$, but also of the actual state $\ket{\psi}$. Indeed, $\rho_{\mathrm{swap}}$ 
is obtained from $\ket{\psi}$ via a local unitary.

Moreover, this bound is a
\textit{lower bound} in two respects: for one, the choice of
${\cal S}$ may not be optimal, i.e. there could be a better guess
for the swap operator; for two, the minimum is taken over a larger
set of correlations than those allowed by quantum physics.

One further clarification is in order: the SDP above~\eqref{fide_Qn} assesses the quality of the state inside the box with respect to the reference state
$\ket{\overline\psi}$ modulo local \emph{isometries}, as opposed to local \emph{unitaries}. Indeed, note that the swapped state $\rho_{swap}$ depends both on
the initial state $\rho$ and on the ancillas $\rho_{A'}$,
$\rho_{B'}$, and it could be the case that the ancillas also
contribute by some amount to this fidelity. For instance, a fully
mixed measured state cannot have a fidelity larger than $1/4$ with
any pure state $\ket{\overline\psi}$; however, pure ancillas can
always reach a fidelity of $1/2$.
This distinction between isometries and unitaries must be taken into account for a proper physical interpretation of the magnitudes derived in this paper.

Note that whenever both the state and measurements are prefectly self-tested, however, the state can be assessed up to local unitaries as well. Indeed, the action of the swap operators can then only be perfect, and therefore the state that is certified only comes from inside the box.

\subsection{General construction}

Here we present a constructive approach to the SWAP method, which
is applicable (though not guaranteed to be optimal) to general Bell inequalities and Bell-type scenarios involving an arbitrary number of parties, inputs and outputs.

\subsubsection{The mathematical guess and convergence conditions for the SWAP method}

Self-testing requires postulating an initial \emph{mathematical
guess} ($\ket{\overline{\psi}}, \{\overline{E}^x_a,
\overline{F}^y_b\}$) on the physics
behind a Bell experiment. The self-testing procedure then assesses
whether the guess is (close to) correct or not.

For instance, in a situation in which we wish to estimate device-independently how close a state prepared in a lab is to the one that we intended to produce, and how close the measurements performed are to the ones we wished to implement, the guessed states/measurements are simply given by the ones we intended to produce/perform.
If, however, we just have access to some distributions
$p(a,b|x,y)$, close to the boundary of the set of quantum
correlations, and we wish to guess the state and measurements
involved, the correlations $p(a,b|x,y)$ must violate some Bell
inequality ${\cal B}$ nearly maximally. Hence, we can apply the
heuristics described in~\cite{I3322_PV} to determine the quantum
state and measurement operators which maximize ${\cal B}$, and
take that to be our mathematical guess. Note that as long as robust self-testing is possible
the guess does not need to be exact, i.e., it is enough that
$\overline p(a,b|x,y) =
\bra{\overline{\psi}}\overline{E}^x_a\otimes
\overline{F}^y_b\ket{\overline{\psi}} \approx p(a,b|x,y)$.

Let us now discuss the conditions under which our method will certify that $p(a,b|x,y)$ self-tests the mathematical guess ($\ket{\overline{\psi}}, \{\overline{E}^x_a,
\overline{F}^y_b\}$) in a robust way. Following the notation of \cite{tsirel_prob}, we say that $p(a,b|x,y)$ is a `relativistic' quantum distribution if there exist a state $\ket{\psi}\in \H$ and projection operators $\mathcal{E}^x_a,\mathcal{F}^y_b\in B(\H)$, with $\sum_a\mathcal{E}^x_a=\sum_b\mathcal{F}^y_b=\id$, and $[\mathcal{E}^x_a,\mathcal{F}^y_b]=0$, for all $x,y,a,b$ such that $p(a,b|x,y)=\bra{\psi}\mathcal{E}^x_{a}\mathcal{F}^y_{b}\ket{\psi}$. Note that the definition of relativistic quantum distributions follows from a relaxation of the tensor structure of Alice and Bob's projection operators. Now, given $p(a,b|x,y)$, consider the following three conditions:

\begin{enumerate}
\item\label{cond1}
The mathematical guess is finite dimensional ($d<\infty$).

\item\label{cond2}
For any sequence of `relativistic' quantum realization $(\ket{\psi_N},\{\mathcal{E}_{a,N}^x, \mathcal{F}_{b,N}^y\})_N$ with $\lim_{N\to\infty}\bra{\psi_N}\mathcal{E}_{a,N}^x\mathcal{F}_{b,N}^y\ket{\psi_N}=\overline{p}(a,b|x,y)$, and any polynomial $Q$ of Alice's and Bob's measurement operators, the identity 
\be
\begin{split}
\lim_{N\to\infty}&\bra{\psi_N}Q(\{\mathcal{E}^x_{a,N},\mathcal{F}^y_{b,N}\})\ket{\psi_N}\\
&=\bra{\overline{\psi}}Q(\{\overline{E}^x_a\otimes \id,\id\otimes \overline{F}^y_b\})\ket{\overline{\psi}},
\label{cond_st}
\end{split}
\ee holds.

\item\label{cond3}
Neither Alice's nor Bob's operators can be simultaneously block-diagonalized, i.e.
\be
\overline{E}^x_a\not=\oplus_k \overline{E}^x_a(k), \text{ and } \overline{F}^y_b\not=\oplus_\ell \overline{F}^y_b(\ell).
\label{blocks}
\ee

\end{enumerate}

If the above conditions apply, then our method will return a sequence of bounds on the desired
property (e.g.: the fidelity with respect to a reference state),
which will converge to the optimal value as the experimental data
$p(a,b|x,y)$ approach $\overline{p}(a,b|x,y)$. This follows from the explicit construction of the SWAP operator given in the next section and from the convergence of the SDP hierarchies for non-commutative polynomial optimization \cite{siam}.

Actually it is generally enough that Eq. (\ref{cond_st}) hold for polynomials Q of bounded degree. For instance, in the previous section, the expression for the singlet fidelity just involved monomials of degree smaller than or equal to 8. Note also that not satisfying condition~\ref{cond2} might prohibit convergence of our method, and hence perfect self-testing of the desired property. However, any bound produced by our method is valid regardless of this condition.

On the other hand, if condition~\ref{cond3} is dropped, then either $\{\overline{E}^x_a\}$ or $\{\overline{F}^y_b\}$ can be expressed as direct sums of elements of the form $\id_{d_k}\otimes \overline{R}_k$, where $\overline{R}_k$ denotes an irreducible representation and $d_k$ is the degeneracy of that representation. 
In that case, convergence is not guaranteed in general, but it can still occur provided that the magnitudes we try to estimate only concern the non-trivial parts of such a block structure. 

For example, let Alice's operators be the direct sum of two non-equivalent irreducible blocks of same dimension, and let $\ket{\overline{\psi}}=\ket{\overline{\psi}_1} \oplus \ket{\overline{\psi}_2}=\ket{0}\otimes\ket{\overline{\psi}_1} + \ket{1}\otimes\ket{\overline{\psi}_2}$ be a decomposition of the state of our mathematical guess on these two blocks. Note that the mixed state $\proj{0}\otimes\proj{\overline{\psi}_1}+\proj{1}\otimes\proj{\overline{\psi}_2}$ would give rise to the same statistics $\overline{p}(a,b|x,y)$. Hence our bounds for the fidelity $\bra{\overline{\psi}}\rho\ket{\overline{\psi}}$ 
do not converge to $1$. However, one can show that the `block fidelity' 
\be
\frac{\bra{0}\bra{\overline{\psi}_1}\rho\ket{0}\ket{\overline{\psi}_1}}{|\braket{\overline{\psi}_1}{\overline{\psi}_1}|^2}+\frac{\bra{1}\bra{\overline{\psi}_2}\rho\ket{1}\ket{\overline{\psi}_2}}{|\braket{\overline{\psi}_2}{\overline{\psi}_2}|^2}
\ee
still converges~\footnote{To see this, consider the following swap operator: $S=\proj{0}\otimes S_1+\proj{1}\otimes S_2$, where $S_1$ and $S_2$ are swaps between each of the blocks and an external register. Let also $T$ be a CNOT that copies the information about which block is occupied into an additional flag register. Applying both $S$ and $T$ to the case in which the measured state is either $\ket{\overline{\psi}_1} \oplus \ket{\overline{\psi}_2}$ or $\proj{0}\otimes\proj{\overline{\psi}_1}+\proj{1}\otimes\proj{\overline{\psi}_2}$ results in external registers being left in the state $\rho=\proj{0}\otimes\proj{\overline{\psi}_1}+\proj{1}\otimes\proj{\overline{\psi}_2}$, which has a block fidelity of 1. The fact that both $S$ and $T$ follow the block structure guarantees, by the Artin-Wedenburn lemma~\cite{takesaki} that they can be obtained as polynomials of the measurement operators, as required.}.

Note that such block fidelity allows for partial self-testing of non-extremal correlations. Thus, it could be used to demonstrate, for instance, that a rank-2 mixed state belongs to the space spanned by its two eigenvectors, even thought this mixed state itself cannot be perfectly self-tested.

For our method to achieve perfect self-testing, we require that
the mathematical guess be finite dimensional. Moreover, we require
that the distribution $\overline{p}(a,b|x,y)$ generated by the
finite-dimensional model ($\ket{\overline{\psi}},
\{\overline{E}^x_a, \overline{F}^y_b\}\subset B(\C^d\otimes\C^d)$)
be such that, for any sequence of quantum distributions
$(p_N(a,b|x,y)=\bra{\psi_N}E^x_{a,N}\otimes
F^y_{b,N}\ket{\psi_N})_N$, with
$\lim_{N\to\infty}p_N(a,b|x,y)=\overline{p}(a,b|x,y)$, there exist
isometries $(W_N)_N$ satisfying \be
\begin{split}
\lim_{N\to\infty}W_NP(\{&E^x_{a,N}\})\otimes Q(\{F^y_{b,N}\})\ket{\psi_N}\\
&=P(\{\overline{E}^x_a\})\otimes Q(\{\overline{F}^y_b\})\ket{\overline{\psi}}\otimes\ket{junk},
\label{cond_st}
\end{split}
\ee for any pair of polynomials $P$ and $Q$ of Alice's and Bob's
measurement operators. Note that, if we further demand the isometries $(W_N)_N$ to be local, this is a strengthening of the
self-testing conditions (\ref{ststate}), (\ref{stmeas}).

Then, under the assumption that Kirchberg's conjecture is true
\cite{ozawa} (i.e., that $({\cal Q}^n)_n$ converges to ${\cal
Q}$), our method will return a sequence of bounds on the desired
property (e.g.: the fidelity with respect to a reference state),
which will converge to the optimal value as the experimental data
$p_N(a,b|x,y)$ approach $p(a,b|x,y)$.

Not satisfying condition~\eqref{cond_st} might prohibit perfect
self-testing of the desired property by our method. However, any
bound it produces is valid regardless of this condition.


\subsubsection{Construction of a unitary swap operator and SDP}
\label{generalSwap}

Since the swaps are local, let us focus again on the construction of the swap operator ${\cal S}_{AA'}$ on Alice's side and omit the subscripts unless they are required. If both $A$ and $A'$ are qudits, an expression for the swap operator is
\ba
{\cal S}_{AA'}&=&TUVU
\label{swapd}
\ea
with
\begin{eqnarray}
&T=\id\otimes \sum_{k=0}^{d-1}\ket{-k}\bra{k},\nonumber\\
&U=\sum_{k=0}^{d-1}P^k\otimes \ketbra{k}{k},\nonumber\\
&V=\sum_{k=0}^{d-1} \ketbra{k}{k}\otimes P^{-k},
\label{uvas}
\end{eqnarray}
where
\begin{equation}\label{Peq}
P=\sum_{k=0}^{d-1}\ket{k+1}\bra{k},
\end{equation}
and additions inside kets are modulo $d$. As before, the idea of the construction consists in mimicking these operators.

We can assume that the algebra generated by the $\{\overline{E}^x_a\}$ is irreducible, i.e., that condition (\ref{blocks}) holds (the case of many irreps can be treated similarly). By the Artin-Wedenburn theorem \cite{takesaki}, any matrix in $\C^d\times \C^d$ on Alice's side is thus an element of the algebra generated by the $\{\overline{E}^x_a\}$. In particular, the operator $P$ in Eq.~\eqref{Peq} can be expressed as a linear combination $P(\overline{E}^x_a)$ of products of Alice's projector operators.


However, contrary to the case of qubits, if in this expression the guesses $\{\overline{E}^x_a\}$ are replaced by arbitrary measurement operators $\{E^x_a\}$, the resulting operator $P(E^x_a)$ needs not be unitary in general. Still, by the polar decomposition \cite{takesaki}, there always exist a unitary~\footnote{Technically, there always exist \emph{an isometry} with the said property. However, any isometry $V\in B(\H)$ in infinite dimensions can be viewed as a unitary operator in $\H\otimes \C^2$. Indeed, let $V^\dagger V=\id$ and define $U=(\id-VV^\dagger)\otimes \ket{0}\bra{1}+V^\dagger\otimes \ket{1}\bra{1}+ V\otimes \ket{0}\bra{0}$. Then $UU^\dagger=U^\dagger U=\id$, and $U\ket{\psi}\ket{0}=(V\ket{\psi})\ket{0}$. At the level of the moment matrices, we can thus assume that such isometries are unitaries.} $\hat{P}$ such that
\ba
\hat{P}^\dagger P(E^x_a)\geq 0.
\label{casimiro}
\ea
Moreover, it is guaranteed that $\hat{P}=P(E^x_a)$ whenever the r.h.s. operator is itself unitary.

Similarly, the projectors $\{\proj{k}\}_{k=0}^{d-1}$ on system $A$ in Eq.~\eqref{uvas} can be replaced by $\{E^0_k\}_{k}$, provided that there are $d$ such projectors and that $\{\overline{E}^0_k\}_{k=0}^{d-1}$ are rank-1. In case one or several of the projectors in Alice's measurement model are degenerate, then we must ``break'' the degeneracy via the addition of new non-commuting variables. For instance, suppose that $\overline{E}^0_k$ has rank $n_k$. Then we must find a self-adjoint element $X_k(\overline{E}^x_a)$ of Alice's algebra of observables such that $\overline{E}^0_kX_k(\overline{E}^x_a)\overline{E}^0_k=\sum_{s=1}^{n_k}\lambda_{k,s}\proj{k_s}$ has $n_k$ different eigenvalues $\lambda_{k,1}>\lambda_{k,2}>...>\lambda_{k,n_k}$. Again, this is always possible by virtue of the Artin-Wedenburn theorem \cite{takesaki}. Now we introduce $n$ new non-commuting variables $\{E^0_{k,j}\}_{j=1}^{n_k}$, which will play the role of $\{\proj{k_j}\}_{j=1}^{n_k}$. These variables must satisfy:
\be
\begin{split}
&E^0_{k,j}E^0_{k,l}=\delta_{j,l}E^0_{k,j}, \sum_{j=1}^{n_k}E^0_{k,j}=E^0_k, [E^0_{k,j},E^0_kX_kE^0_k]=0,\\
&\frac{1}{2}(\lambda_{k,j}+\lambda_{k,j+1})E^0_{k,j}\geq E^0_kX_k(E^x_a)E^0_{k,j}, j=1,...,n_k-1\\
&E^0_kX_k(E^x_a)E^0_{k,j}\geq \frac{1}{2}(\lambda_{k,j-1}+\lambda_{k,j})E^0_{k,j}, j=2,...,n_k.
\end{split}
\label{spectrum}
\ee
\noindent As with $\hat{P}$, the existence of the projectors $\{E^0_{k,j}\}$ does not impose extra conditions, and can always be taken for granted.

Now we can collect all the elements of the construction of the swap operator on Alice's side:
\begin{enumerate}
\item Guess the operators $\overline{E}^x_a$.
\item Construct $P$ given in \eqref{Peq} as linear combinations of products of the $\overline{E}^x_a$. Similarly, for each degenerate projector $E^0_k$, find $X_k(\overline{E}^x_a)$ such that $\overline{E}^0_kX_k(\overline{E}^x_a)\overline{E}^0_k$ is non-degenerate in the support of $\overline{E}^0_k$.
\item Formally replace $\overline{E}^x_a$ by the unknown ${E}^x_a$ in those expressions to obtain the expressions of $P(E^x_a)$ and $X_k(E^x_a)$ (if needed).
\item Define the swap operator as \eqref{swapd}, in which $U$ and $V$ are given by the expressions \eqref{uvas} with $\proj{k}$ replaced by $E^0_{k}$ or $E^0_{k,j}$ and $P$ replaced by $\hat{P}$ (on the original system). These otherwise undefined operators are constrained in terms of the ${E}^x_a$ by \eqref{casimiro} and \eqref{spectrum}.
\end{enumerate}

The inclusion of $\hat{P}_A, \hat{P}_B,E^0_{k,j}$ in the moment matrix, together with the extra semidefinite constraints \eqref{casimiro}, \eqref{spectrum} is known as the technique of \emph{localizing matrices} \cite{siam}. We review it here.

Let $g=\sum_u g_u u$ be a polynomial of Alice and Bob's measurement operators (with the $u$'s being operator products), and let $\cc$ be a moment vector. Then, the \emph{localizing matrix} $\Gamma^{S}(g,\cc)$ is a matrix whose rows and columns are numbered by elements of the set of products $S$, and such that $\Gamma^{S}_{s,t}(g,\cc)=\sum_{u} g_u\cc_{s^\dagger u t}$. It can be verified that, if $\cc$ is such that it admits a quantum representation where the polynomial $g$ is a non-negative operator, then $\Gamma^{S}_{s,t}(g,\cc)$ must be positive semidefinite.

In our scenario, we must guarantee that the optimization is done over all quantum representations such that (\ref{casimiro}), (\ref{spectrum}) hold. Such constraints hence translate to:
\be
\Gamma^{S'}(\hat{P}_A^\dagger P_A(E^x_a),\cc)\geq 0\ ,\ \  \Gamma^{S'}(\hat{P}_B^\dagger P_B(F^y_b),\cc)\geq 0,
\ee
plus the constraints associated to conditions (\ref{spectrum}). Here $S'$ is chosen as big as possible, but such that all entries of the localizing matrices can be written as linear combinations of moment vectors defined over $SS^\dagger$. Note that requiring $\Gamma^{S'}(\hat{P}_A^\dagger P_A(E^x_a))$ to be positive also implies that it must be hermitian.

The semidefinite program to lower bound the fidelity of the swapped state with respect to the reference state $\ket{\overline{\psi}}$ is then:

\begin{align}
	f^S=\min \;\; &\bracket{\overline{\psi}}{\rho_{\mathrm{swap}}(\cc)}{\overline{\psi}} \nonumber\\
	\textrm{s.t.}\hspace{0.2cm} & \Gamma^S(\cc) \geq 0, \nonumber\\
			    & \cc_{E^x_aF^y_b}=p(a,b|x,y) \nonumber\\
			    &\Gamma^{S'}(\hat{P}_A^\dagger P_A(E^x_a),\cc)\geq 0, \nonumber\\
			    &\Gamma^{S'}(\hat{P}_B^\dagger P_B(F^y_b),\cc)\geq 0,
			    \label{cosica}
\end{align}

\noindent plus extra constraints in case a subset of the $\overline{E}^0_k$'s ($\overline{F}^0_k$'s) is degenerate.

Note that whenever the reference state and measurements are chosen real, it is sufficient to perform this optimization over real SDP matrices, because the objective function is a combination of moments with real coefficients.

\subsection{Finite-size fluctuations, beyond i.i.d.}
\label{non_iid}

All the previous discussion implicitly assumed that the behavior
of the devices is the same in each run and is uncorrelated among
the runs, that is the \textit{i.i.d. assumption}. Moreover, we
presented the case for infinitely many runs of the experiment,
such that $p(a,b|x,y)$ can be estimated exactly. In this
paragraph, we remove both assumptions, by presenting a finite-size
analysis inspired by \cite{Zhang11}. As one of the outcomes, we
prove that the asymptotic bounds can be computed under the i.i.d.
assumption without loss of generality.

Suppose that Alice and Bob have sequentially distributed pairs of
black boxes. We allow for the possibility that different pairs of
boxes exhibit different statistics, which can, in turn, depend on
Alice and Bob's past measurement history. Now, let $g$ be a
function of the underlying state and measurement operators in each
realization for which the SWAP tool, or any other method,
establishes that $g(\ket{\psi},E^x_a,F^y_b) > g^*$, for some $g^*$,
whenever the violation of a specific Bell inequality $\B$ via i.i.d. pairs is greater than $V_0$.
Then, given a Bell violation $V>V_0$ obtained with non-i.i.d. boxes, we wish to disprove:


\vspace{5pt}

\noindent \textbf{Hypothesis $\Phi$}

\emph{All the distributed pairs contain quantum states and operators $(\ket{\psi},E^x_a,F^y_b)$ such that $g(\ket{\psi},E^x_a,F^y_b)\leq g^*$.}

\vspace{5pt}

To do this, the idea is to define a statistical parameter $T$
which both parties can estimate during the course of the
experiment and such that $P(T>1/\delta|\Phi)<\delta$. If the
observed value $t$ is such that $t>1/\delta_0$ for some threshold
$\delta_0$, the parties can conclude that hypothesis $\Phi$ is not
likely to be true. Let us construct this parameter $T$.

Let
$\Psi\equiv(\overline{\psi},\overline{E}^x_a,\overline{F}^y_b)$ be
a particular quantum model with $\B$-violation $V>V_0$. Under the
assumption that Alice and Bob can choose their measurement
settings $x,y$ randomly and independently of their boxes, any Bell
inequality $\B$ can be written as $\langle B(a,b,x,y)\rangle\leq
V_0$, with $B(a,b,x,y)$ being an arbitrary real function of the
inputs and outputs of the problem that will depend on Alice and
Bob's distribution $p(x)p(y)$ of the inputs.

Let $|B(a,b,x,y)|\leq K$ for all inputs and outputs. Following the
lines of \cite{Zhang11}, we define the normalized form of
$B(a,b,x,y)$ as $\tilde{B}(a,b,x,y)\equiv
\frac{B(a,b,x,y)+K}{V_0+K}$. Clearly, $\langle
\tilde{B}(a,b,x,y)\rangle_\Psi> 1$, and $\tilde{B}(a,b,x,y)\geq 0$
for all $x,y,a,b$. Also, $\langle \tilde{B}(a,b,x,y)\rangle\leq 1$
for any pair of boxes satisfying hypothesis $\Phi$.

Next, choose $0<\epsilon<1$ such that \be R(a,b,x,y)\equiv
(1-\epsilon)+\epsilon \tilde{B}(a,b,x,y) \ee satisfies \be \langle
\log[R(a,b,x,y)]\rangle_\Psi>0. \label{log_posi} \ee That such an
$\epsilon$ exists follows from the observation that, for
$\epsilon\ll 1$, \be \langle\log(1-\epsilon+\epsilon
\tilde{B}(a,b,x,y))\rangle_\Psi\approx \epsilon
\langle(\tilde{B}(a,b,x,y)-1)\rangle_\Psi>0. \ee Note that, by
construction, $\langle R(a,b,x,y)\rangle\leq 1$ under hypothesis
$\Phi$.

Now, suppose that Alice and Bob conduct the Bell experiment $n$
times, choosing their inputs $x,y$ with probability $p(x)p(y)$
each time, thus obtaining the experimental data
$\{a_k,b_k,x_k,y_k\}_{k=1}^n$. Define the positive random variable
$T\equiv \prod_{k=1}^n R_k$, with $R_k\equiv R(a_k,b_k,x_k,y_k)$.
Under hypothesis $\Phi$, it can be seen that $\langle T\rangle\leq
1$ \cite{Zhang11}, and so, by Markov's inequality, $P(T\geq
\delta)\leq 1/\delta$.

However, in the event that Alice and Bob
are actually being distributed $n$ independent copies of box
$\Psi$, by the central limit theorem, the random variable
$X\equiv\log(T)=\sum_{k=1}^n \log(R_k)$ is expected to take values
in the range $n\langle\log(R)\rangle_\Psi\pm O(\sqrt{n})$. From
Eq. (\ref{log_posi}), we thus have that, with very high
probability, $T$ will grow exponentially with $n$. In a few
experiments, Alice and Bob will hence observe a ridiculously high
value of $T$, and therefore conclude that hypothesis $\Phi$ must
be abandoned.

A rough estimate on the probability of (wrongly) accepting
hypothesis $\Phi$ when $n$ independent copies of $\Psi$ are
actually distributed can be established via Chebyshev's
inequality, which states that, for any random variable $Z$,
$P(|Z-\langle Z\rangle|\geq\epsilon)\leq \frac{\langle
Z^2\rangle-\langle Z\rangle^2}{\epsilon^2}$. Let $\delta_0>0$ define the criterion used to
reject hypothesis $\Phi$, i.e., Alice and Bob will reject $\Phi$
iff $T>1/\delta_0$. Suppose also that $n$ is large enough to
guarantee that $\langle X\rangle_\Psi=n\langle
\log(R)\rangle_\Psi\geq \log(\delta_0^{-1})$. Then, the
probability $P\left(T\leq\delta_0^{-1}\right)$ that $\Phi$ is
accepted satisfies
\begin{eqnarray}
P\left(T\leq\delta_0^{-1}\right)&=& P\left(\langle X\rangle-X\geq\langle X\rangle-\log(\delta_0^{-1})\right)\nonumber\\
&\leq& P\left(|\langle X\rangle-X|\geq|\langle X\rangle-\log(\delta_0^{-1})|\right)\nonumber\\
&\leq& \frac{\langle\log^2(R)\rangle_{\Psi}-\langle\log(R)\rangle_{\Psi}^2}{n\left(\langle \log(R)\rangle_\Psi-\frac{\log(\delta_0^{-1})}{n}\right)^2},
\end{eqnarray}
which tends to zero as $O(1/n)$.

In order to reject hypotheses such as ``the singlet fidelity of
the state inside the boxes is smaller than $f^*$ for each
realization'', it is thus enough to estimate the maximal Bell
violation $\B$ compatible with fidelity $f^*$ in the i.i.d. case.

The method so far described, though, is based on the estimation of
the violation of a \emph{single} Bell inequality. One could ask
what happens, then, when we consider additional parameters in our
i.i.d. analysis, such as the whole probability distribution
$p(a,b|x,y)$. As we will see below, the minimum singlet fidelity
for a given CHSH violation substantially increases when the
experimental distribution is isotropic, see Section
\ref{CHSH_snglt} for details.

In fact one can show, following an argument similar to that presented in~\cite{olmo, moreRandomness}, that such improved bounds can be demonstrated by monitoring some other Bell inequality $\B'$, described by the dual of our SDP program~\eqref{cosica}. One can thus also consider these improved bounds in presence of finite statistics by applying the analysis above to the new Bell inequality $\B'$\footnote{In practice, the whole distribution $p(a,b|x,y)$ is generally not accessible. Rather, frequencies $f(a,b|x,y)$ can be observed. One may thus be tempted to use these frequencies in place of the distribution $p$ itself in order to guess the new Bell inequality $\B'$. These frequencies, however, typically don't satisfy the no-signaling condition, and thus don't belong to any level of the NPA relaxation. Therefore, they cannot be set as constraints in our SDP program. One should thus not apply the SDP to the observed frequencies directly, but rather to some quantum point $p$ chosen reasonably close to $f$ (if no quantum point is statistically close to $f$, then the validity of the experiment may be questionned). The bound obtained with the inequality found in this way will be valid independently of how the frequencies were approximated.}.

%
%
%

Note that ruling out hypothesis $\Phi$ guarantees that at least one of the pairs is of the desired kind, i.e. has $g>g^*$. In a case where $n$ runs have taken place, one might want to guarantee that at least a significant fraction of these $n$ runs satisfy $g>g^*$. We leave this general question open for further study. In the special case where the finite statistics experiment was realized with i.i.d. boxes, however, our argument already guarantees that with large probability all boxes are of good quality.

In view of these reflections, along the rest of the article we
will always work in the asymptotic case of infinitely many runs
and the i.i.d. behavior of the boxes will be taken for granted.

\section{More robust bounds for known self-testing scenarios}
\label{more_robust}

The remainder of the paper is devoted to discussing several
explicit examples of self-testing. We start with examples that are
already known in the literature to have robust self-testing, and
show how our method obtains much stronger bounds.

\subsection{Singlet fidelity from CHSH}
\label{CHSH_snglt}

The first reported example of self-testing, though it was not
called that way, is the fact that the maximal violation of the
CHSH inequality self-tests the maximally entangled state of two
qubits \cite{sw87,pr92}. The CHSH expression is \ba
CHSH=\langle A_0B_0 \rangle + \langle A_0B_1\rangle + \langle A_1B_0\rangle -\langle A_1B_1\rangle \,. \ea For simplicity of
notation, we work with a guessed ideal case in which both Alice's
and Bob's observables are the same, i.e. \ba
\overline{A}_0=\overline{B}_0=\sigma_z&,&
\overline{A}_1=\overline{B}_1=\sigma_x\,.\label{XZ} \ea This
requires writing the maximally entangled two-qubit state as \ba
\ket{\overline{\psi}}&=&\cos\left(\frac{\pi}{8}\right)\ket{\Phi^-}+\sin\left(\frac{\pi}{8}\right)\ket{\Psi^+}\,.
\ea Having set this, we choose the initial states of the ancilla
qubits to be $\ket{0}_{A'}$ and $\ket{0}_{B'}$, then go through
the procedure described in Section~\ref{sec:exqubit} without
modification to obtain after some algebra \ba
F&=&\frac{1}{2}\,+\,\frac{1}{2\sqrt{2}}\,CHSH-\frac{1}{8}\ave{[A_0,A_1][B_0,B_1]}\nonumber\\
&&+\frac{1}{8\sqrt{2}}\Big( 3\ave{A_1B_1}-2\ave{A_0B_1}-2\ave{A_1B_0}\nonumber\\
&&+\ave{A_0A_1A_0(B_1-2B_0)}+\ave{(A_1-2A_0)B_0B_1B_0}\nonumber\\
&&-\ave{A_0A_1A_0B_0B_1B_0}\Big)\,.\label{Fchsh} \ea It is
presently not difficult to check, with the tools used in the proof
of the Tsirelson bound, that $F=1$ indeed if
$CHSH=2\sqrt{2}$ \footnote{Looking for the eigenvalues of
the square of the CHSH operator, one finds that $CHSH=2\sqrt{2}$ can
only be obtained if $\ave{[A_0,A_1][B_0,B_1]}=-4$. Together with
the unitarity of the operators, this implies $A_0A_1=-A_1A_0$ i.e.
$A_0A_1A_0=-A_1$, and the same for Bob. By replacing all these
results in \eqref{Fchsh}, one finds $F=1$.}. Our goal however is
to go beyond the ideal case and obtain a lower bound on $F$ via
SDP.

\begin{figure}[htbp!]
    \begin{center}
    \includegraphics[width=0.48\textwidth]{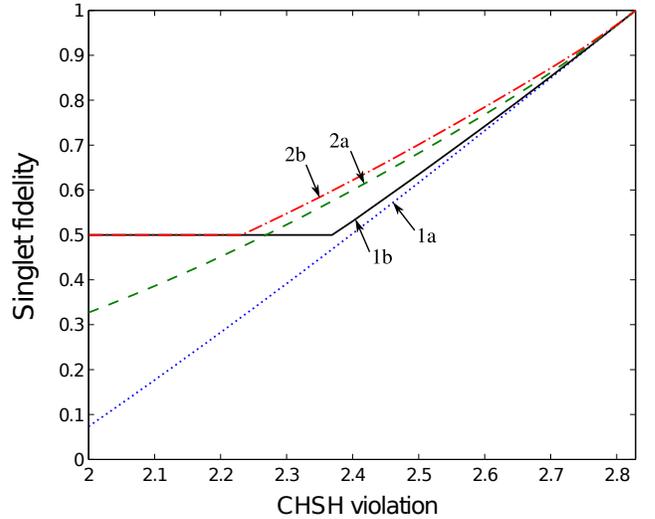}
    \caption{Minimal singlet fidelity as a function of CHSH violation. Curves 1(2) correspond to lowerbounds on the fidelity for generic(isotropic) boxes. Curves a(b) were obtained with the un-modified(improved) swap operator.}
    \label{fid_CHSH}
    \label{fid_CHSH_better}
\end{center}
\end{figure}

The constraint can be the value of $CHSH$ alone, or all the
statistics $p(a,b|x,y)$ collected during the experiment. The
second case takes advantage of more information than the first one
and thus provides a better bound. This is illustrated by curves 1a
and 2a in Figure \ref{fid_CHSH}.

As mentioned previously, there is no guarantee that the swap we
used [defined by Eq. (\ref{uvu}, \ref{uvqubit})] is optimal. In fact,
we have found better bounds by tweaking the standard method
in two ways. Firstly, we consider swap operators which
\emph{depend on the Bell violation}. The first dependence we
consider consists in using the dichotomic operator $A_2$ satisfying

\be
A_2\left(\cos(\zeta_A(V))A_0+\sin(\zeta_A(V))A_1\right)\geq 0,
\ee

\noindent where $\zeta_A(V)$ is an arbitrary function of the CHSH
violation $V$, as Alice's logical `NOT' operator.
For $V=2\sqrt{2}$ choosing $\zeta_A(V)=\pi/2$ allows
one to recover $A_2=A_1$ so that the NOT operator returns the
correct SWAP in the case of maximal CHSH violation. Similarly, we
can parametrize an auxiliary operator for Bob by a function
$\zeta_B(V)$.

The intuition behind this swap ansatz is that, even though Alice
and Bob are preparing the maximally entangled state, their
measurement devices are somehow ``tilted'', thus explaining why
they do not achieve the optimal CHSH value. The resulting fidelity
will have the same expression as in Eq.~(\ref{Fchsh}), but with
$A_1,B_1$ replaced by $A_2,B_2$.

We introduce a second parameter in the swap operator by considering combinations of $UVU$ with the identity, i.e. we define
\be
\mathcal{S}_{AA'} = \cos \xi_A\, \openone + i \sin \xi_A\, UVU
\ee
and similarly for Bob. For every value of $\xi_A$, $\mathcal{S}_{AA'}$ is still a unitary operator, and setting $\xi_A=\pi/2$ recovers the previous choice of swap operator.

The second tweak that we apply on the standard SWAP method takes advantage of the fact that since we
are only interested in certifying the quality of the measured
state up to local isometries, it is sufficient to lowerbound its
fidelity with respect to any reference state of the form \be
\ket{\psi_{ME}} = W_A\otimes W_B \ket{\overline \psi}, \ee where
$W_A$ and $W_B$ are arbitrary single qubit unitaries.

Altogether, we thus looked for parameters $\zeta_A$, $\zeta_B$, $\xi_A$, $\xi_B$ and
unitaries $W_A$, $W_B$ that could improve the bounds 1a and 2a in
Figure \ref{fid_CHSH}. The result of this optimization is shown in
the same figure with curves 1b and 2b.

The new curves display a different behavior in two regions. When $CHSH\gtrsim 2.37$ (or $CHSH \gtrsim 2.23$ in the isotropic case), an appropriate choice of $\zeta_A$, $\zeta_B$, $W_A$, $W_B$ together with $\xi_A=\xi_B=\pi/2$ results in an improved bound on the fidelity compared to the standard SWAP procedure. For lower CHSH violations, however, the best choice of swap operator is obtained with $\xi_A=\xi_B=0$, resulting in a singlet fidelity of $1/2$. This value is thus obtained by comparing the ancilla rather than the state inside the box to the singlet (or equivalently by considering the local isometry that maps every local Hilbert space to the pure state $\ket{0}$). It thus corresponds to a trivial singlet fidelity, that can be certified up to local isometries regardless of the measured state. Clearly, any self-testing performed up to local isometries admits a trivial fidelity bound of this kind.

One can check that the curve 1a can be saturated by some two-qubit states and measurements when the standard swap~(\ref{uvu}, \ref{uvqubit}) and reference state $\ket{\overline \psi}$ are used. This bound is thus tight under this condition. Therefore, a modification of the SWAP procedure was necessary to improve this bound. It would be interesting to see if the improved bounds can be improved further or if they are close to optimal.

The fact that the bounds obtained in the isotropic and generic
case differ in the first region suggests that the CHSH inequality is not optimal to
bound the singlet fidelity of singlets subject to uniform noise.
Upon examination of the dual of our SDP, we find that a better
Bell expression to consider is of the form
\begin{equation}
I_{\alpha,\beta}=\alpha \langle A_0 B_0 \rangle + \langle A_0 B_1 \rangle + \langle A_1 B_0 \rangle - \beta \langle A_1 B_1 \rangle\,,
\end{equation}
where $\alpha,\beta>0$ are parameters which depend on the quality
of the statistics.

Note that an inequality of the same form, but with different values for the coefficients $\alpha$, $\beta$, was also shown to be relevant for randomness certification in presence of a Werner state~\cite{olmo, moreMore}.

\vspace{10pt}

\noindent\emph{Comparison with previous bounds on the singlet fidelity from the CHSH violation}
\vspace{5pt}

Previous to this work, most self-testing bounds were given in
terms of the norm of the difference between the desired state
$\ket{\overline{\psi}}_{A'B'}\otimes\ket{\textrm{junk}}_{AB}$ and
the actual one after the isometry $\ket{\psi}_{A'B'AB}$, i.e. in
the form \ba
||\ket{\psi}-\ket{\overline{\psi}}\otimes\ket{\textrm{junk}}||&\leq&t\,.
\ea As a consequence, \ban t^2&\geq&2-2s \ean where the scalar
product $s=\braket{\psi}{\overline{\psi}\otimes\textrm{junk}}$ can
be taken real and positive since everything is defined up to a
local unitary, so in particular it is possible to add a suitable
global phase.

In this text we work with the singlet fidelity evaluated on the
swapped state, i.e.
$F=\bracket{\overline{\psi}}{\rho_\text{swap}}{\overline{\psi}}$.
In Appendix \ref{relation}, we show that both quantities are
related through
\be
s^2\geq 2F-1.
\label{rela}
\ee

For a CHSH violation of $2\sqrt{2}-\epsilon$, curve $1b$ in
Figure~\ref{fid_CHSH_better} gives a lower bound of approximately
$F\gtrsim 1-1.1\epsilon$. We thus have that
\begin{eqnarray}\label{eq:2.2eps}
||\ket{\psi}-\ket{\overline{\psi}}\otimes\ket{\textrm{junk}}||^2&=& 2(1-s) \nonumber\\
&\leq& 2-2\sqrt{2F-1}\nonumber\\
&\lesssim& 2.2\epsilon.
\end{eqnarray}

In the work of McKague et al.~\cite{mckague}, the values of $t$
are given explicitly by replacing $\epsilon_1$ and $\epsilon_2$
with the values given in Theorem 2 for CHSH and in Theorem 3 for
Mayers-Yao, which gives:
\begin{eqnarray}
||\ket{\psi}-\ket{\overline{\psi}}\otimes\ket{\textrm{junk}}||^2
&\lesssim& 100\sqrt{\epsilon}
\end{eqnarray}
Note that this bound is not linear in $\epsilon$, and thus seems harder to compare with~\eqref{eq:2.2eps}. However, one can see that this bound quickly becomes trivial. Indeed, already for $\epsilon\simeq 2.2\cdot 10^{-5}$, it yields a distance larger than $2-\sqrt{2}$, the distance between the singlet state and a product state.

In the work of Reichardt et al.~\cite{reichardt}, the bounds are
not explicitly given, but can be reconstructed. The relevant
result for comparison with our work is Lemma 4.2 in the long
version \cite{reichardtlong}, which refers to the single CHSH game
(i.e. before the extension to parallel repetition that constitutes
the main result of their work). With their notations:
\be\label{boundnorm}
\begin{split}
||\ket{&\bar{\psi}}-\ket{\psi^*}\otimes\ket{\psi^{\times}}||^2\\
=&\,\sum_{c,i,j} ||\psi_{cij}||^2\,||\ket{\tilde{\psi}_{cij}}-e^{i\phi_{cij}}\ket{\psi^*}||^2\\
&\stackrel{(*)}{\leq} 72284\,\sum_{c,i,j}
||\psi_{cij}||^2\,\beta_{ij}\,\stackrel{(**)}{\lesssim}
10^5\,\epsilon 
\end{split}
\ee where (*) comes from the end of proof of
Proposition 4.5 and (**) from Eq.~(4.4), using
$72284*\sqrt{2}\approx 10^5$.

We see that our results are orders of magnitude better than the
best previous results in the literature. Note also that no other
results allow us to establish a non-trivial bound for the singlet
fidelity in the best CHSH experiment realized so far
\cite{Christensen13}. Curve $1b$ in Figure~\ref{fid_CHSH_better},
on the other hand, allows one to certify a singlet fidelity
greater than 99\%.

\subsection{Singlet fidelity from Mayers-Yao}
\label{mayers_yao}

The work of Mayers and Yao, where the wording ``self-testing" was
first used, reported another criterion to self-test the maximally
entangled state of two qubits. This time,
$\ket{\overline{\psi}}=\ket{\Phi^+}$, the measurements are the
same as \eqref{XZ} but there is a third setting on Alice's side,
ideally \ba \overline{A}_2&=&\frac{\sigma_z+\sigma_x}{\sqrt{2}}\,.
\ea The ideal Mayers-Yao statistics, which self-tests
$\ket{\Phi^+}$ with $F=1$, are given by \be
\begin{array}{lcl}
\ave{A_0B_0}\,=\,\ave{A_1B_1}&=&1\\
\ave{A_0B_1}\,=\,\ave{A_1B_0}&=&0\\
\ave{A_2B_0}\,=\,\ave{A_2B_1}&=&\frac{1}{\sqrt{2}}\,.
\end{array}\label{MYideal}
\ee
As for the CHSH example, we can just use the swap defined by \eqref{uvu} and \eqref{uvqubit} to obtain
\ba
F&=&\frac{1}{4}(1+\ave{A_0B_0})\,+\,\frac{1}{16}\,\Big(\ave{[A_0,A_1][B_0,B_1]}\nonumber\\
&&+\ave{(A_1-A_0A_1A_0)(B_1-B_0B_1B_0)}\Big)\,.\label{FMY} \ea
Notice that the setting $A_2$ does not appear in the construction
of the swap, nor consequently in $F$; but it fulfills the crucial
role of inducing constraints on higher moments of
$A_0,A_1,B_0,B_1$ in the SDP matrix.

For definiteness, we study the robustness of a mixture of ideal
Mayers-Yao statistics (weight $v$) with white noise (weight
$1-v$): that is, the statistics are given by the correlations
\eqref{MYideal} multiplied each by $v$. These are the correlations
expected from using the prescribed measurements on a Werner state
$\rho_W=v\ketbra{\Phi^+}{\Phi^+}+(1-v)\openone/4$ with visibility
$v$.

\begin{figure}
\includegraphics[width=0.48\textwidth]{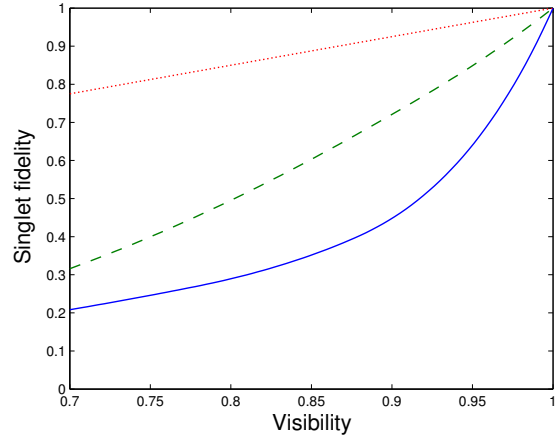}
\caption{Lower bounds on the singlet fidelity of Werner states certified by the Mayers-Yao test (full line). Same bound as certified by the CHSH test is reported here from Figure~\ref{fid_CHSH} for comparison (dashed line). The dotted line is the actual fidelity of the Werner state itself.}
\label{fig:MY}
\end{figure}

Figure~\ref{fig:MY} shows the bound on the fidelity that we are
able to certify from a direct application of our method for different values of $v$. We
compare it to the analogous bound obtained from the CHSH violation
achievable by the state and the actual fidelity of the Werner
state. None of the lower bounds come close to the actual fidelity,
thus suggesting that different measurements are necessary to
optimally bound the singlet fidelity of Werner states in a
device-independent manner.

\subsection{Complementary qubit measurements from CHSH}
\label{measqubit}

Suppose that, rather than verifying that $\ket{\psi}$ is close to
$\ket{\overline{\psi}}$, we are interested in certifying to which
degree Bob's input $y$ is able to switch between the qubit
dichotomic measurements $\sigma_z,\sigma_x$. For that, we envision
a virtual experiment in which we hand Bob a trusted qubit in a
state $\ket{\varphi}$ unknown to him. We let Bob turn on an
interaction between the trusted system and his box and only then
we tell Bob which symbol $y=0,1$ to input. If the resulting
statistics $P(b|y,\ket{\phi})$ satisfy
$P(b|0,\ket{\varphi})=\bra{\varphi}\frac{\id+(-1)^b\sigma_z}{2}\ket{\varphi}$,
$P(b|1,\ket{\varphi})=\bra{\varphi}\frac{\id+(-1)^b\sigma_x}{2}\ket{\varphi}$,
we conclude that Bob's input indeed allows him to switch between
the Pauli measurements $\sigma_z,\sigma_x$.

\begin{figure}[htbp!]
    \begin{center}
    \includegraphics[scale=0.3]{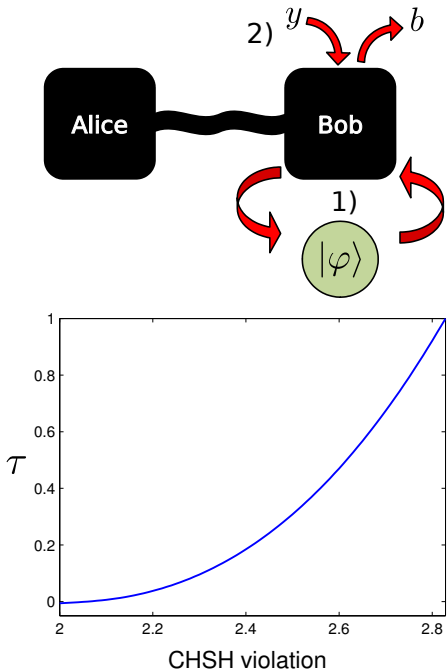}
    \caption{Estimation of Bob's measurements. The protocol works in two steps: 1) We implement a full SWAP of Bob's box and his trusted qubit, that we prepare in state $\ket{\varphi}$. 2) We implement measurement $B_y$ and study the resulting statistics.}
    \label{meas_CHSH}
\end{center}
\end{figure}

Considering that Bob uses the swap operator specified in
Eqs.~(\ref{uvu}, \ref{uvqubit}) as an interaction between the
state $\ket{\varphi}$ and his box, he will observe outcome $b$,
when later performing measurement $y$, with probability
\begin{equation}
P(b|y,\ket{\varphi}) = \tr\left(M^y_b \cal{S} \rho_{AB}\otimes
\ket{\varphi}_{B'}\bra{\varphi} \cal{S}^\dagger\right),
\end{equation}
where $\cal{S}$ is given by Eq.~(\ref{defS}).

To quantify the hypothesis $B_0=\sigma_z,B_1=\sigma_x$, we define the figure of merit
\begin{equation}
\begin{split}
\tau\equiv \,&\frac{1}{2}\left\{P(0|0,\ket{0})+P(1|0,\ket{1})+\right.\\
& \left.+P(0|1,\ket{+})+P(1|1,\ket{-})\right\}-1
\end{split}
\end{equation}
which compares statistics obtained when using different states for
the ancillas. $\tau$ is a number ranging from -1 to +1, and
$\tau=1$ is achievable only in the ideal case.

As before, $P(b|y,\ket{\varphi})$ depends linearly on the moment vector $c$. For example:
\begin{eqnarray}
&P(0|0,\ket{0})=\bra{\psi}\left(\frac{\id+B_0}{2}\right)\ket{\psi}+\nonumber\\
&+\bra{\psi}\left(\frac{\id-B_0}{2}\right)B_1\left(\frac{\id+B_0}{2}\right)B_1\left(\frac{\id-B_0}{2}\right)\ket{\psi}.
\end{eqnarray}

\noindent Consequently, $\tau=\tau(\cc)$, and so a lower bound on
this quantity can be estimated with an SDP for measurements that
are compatible with some Bell violation. The result of this
computation for the case of measurements leading to some CHSH
violation is shown in Figure \ref{meas_CHSH}, where $\tau$ indeed
tends to 1 as the CHSH parameter gets closer to $2\sqrt{2}$.

\subsection{Fidelity of partially entangled two-qubit pure states}
\label{partialqubit}

In Sections \ref{CHSH_snglt}, \ref{mayers_yao}, we have used
quantum nonlocality to certify how close Alice and Bob's state
$\ket{\psi}$ is to a maximally entangled qubit pair. Here, we show
how our method can be used to certify non-maximally entangled
qubit pairs as well. This problem was first considered in
\cite{yang}, where an analytic estimate on the state fidelity was
obtained. Here, using the SWAP concept, we derive a much better
bound.

The scenario is similar to that used above to self-test the
singlet state. The only difference is that we will use a tilted
CHSH inequality of the form
\begin{align}
    \mathcal{B}_\alpha = \alpha \ave{A_0} + \ave{A_0B_0}+\ave{A_0B_1}+\ave{A_1B_0}-\ave{A_1B_1},
\end{align}
where $0\leq\alpha\leq2$. The maximum quantum violation of this inequality \cite{acintilted} is given by $\sqrt{8+2\alpha^2}$ and the corresponding optimal qubit strategy is
\begin{align}
    \overline{A}_0&=\sigma_z, \nonumber\\
    \overline{A}_1&=\sigma_x, \nonumber\\
    \overline{B}_0&= \cos\mu\;\sigma_z+\sin\mu\;\sigma_x, \nonumber\\
    \overline{B}_1&= \cos\mu\;\sigma_z-\sin\mu\;\sigma_x, \nonumber\\
    \ket{\overline{\psi}} &= \cos\theta\ket{00}+\sin\theta\ket{11},\label{optimaltilted}
\end{align}
where $\sin2\theta = \sqrt{ (1-\alpha^2/4)/(1+\alpha^2/4)}$ and
$\tan\mu=\sin2\theta$. Thus, we can use the appropriate inequality
with the corresponding value of $\alpha$ to device independently
certify a reference state of the form
$\ket{\overline{\psi}}=\cos\theta\ket{00}+\sin\theta\ket{11}$. The
details on the construction of the swap operator are similar to
those in Section \ref{CHSH_snglt}, and can be found in Appendix
\ref{tilted_app}.

Figure \ref{tiltedCHSH} shows a plot of the minimal fidelity with
$\ket{\overline\psi}$ for different values of $\alpha$ and
different Bell inequality violations.

\begin{figure}[htbp!]
    \begin{center}
    \includegraphics[width=0.48\textwidth]{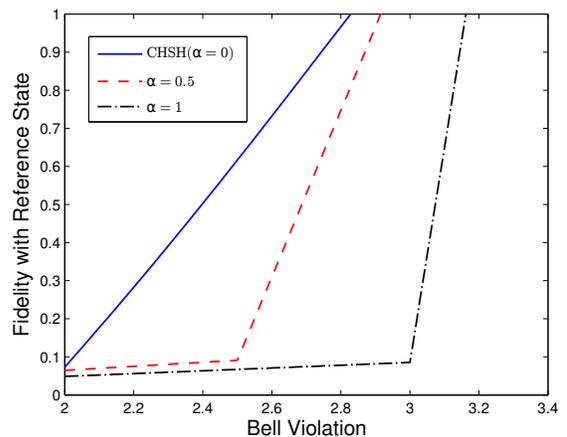}
    \caption{The solid line refers to the standard CHSH case, which obviously recovers the line in Figure \ref{fid_CHSH}. The dashed line refers to the case when $\alpha=0.5$ and the local bound is 2.5, explaining the discontinuity. The dotted-dashed line refers to the case when $\alpha=1$ and the local bound is 3. \label{tiltedCHSH}}
\end{center}
\end{figure}

The curves are approximately linear, between the local bound and
maximum quantum bound. Since the local and quantum bound coincide
at $\alpha=2$, the line gets steeper. This is expected, because,
as $\alpha$ increases, the range of tolerable error gets smaller.
However, it is interesting to note that the robustness is always
close to linear for the three cases.

\section{New robust self-testing scenarios}
\label{new_robust}

\subsection{Fidelity of a pure two-qutrit state from CGLMP}
\label{CGLMP_app}

So far, our studied cases concern scenarios where Alice and Bob
each have two inputs and two outputs, and they want to certify
their state against an entangled pair of qubits. In this section,
we illustrate how to extend these ideas to higher dimensional
scenarios. It is unclear how one can generalize the methods
sketched in \cite{mckague,reichardt,yang} for such a purpose.
However, using the SWAP method above, it is easy to do so.

The relevant Bell inequality for this case is the CGLMP inequality
\cite{cglmp}; it requires two measurement settings on each side,
with three measurement outcomes. The inequality reads:
\begin{align}
    &\mathcal{B}_{\mathrm{CGLMP}}(p) = p(a<b|x=1,y=1)+\nonumber\\
    &+p(a>b|x=0,y=1)+ p(a\geq b|x=1,y=0)+\nonumber\\
    &+p(a<b|x=0,y=0) \geq 1.
\end{align}
The maximum quantum violation is conjectured \cite{conject} and
verified numerically \cite{navascues} to be
$\mathcal{B}_{\mathrm{CGLMP}}(p)=(12-\sqrt{33})/9\approx 0.6950$.
Moreover, it is believed that the maximal quantum violation can
only be achieved with the (non-maximally entangled) state and
measurement operators described in \cite{conject,cglmp}. Here we
will prove this conjecture true.

First of all, let us re-express the state and measurement
of~\cite{conject,cglmp} in real form. This can be achieved by
choosing the first measurements of Alice's and Bob's to be
$\overline{E}_m^0=\overline{F}_m^0=\ketbra{m}{m}$ for $m=0,1,2$,
the second ones
$\overline{E}_m^1=\overline{F}_m^1=\proj{\alpha_m}$ with
\begin{equation}
\ket{\alpha_m} = \dfrac{1}{3} \left( 2\ket{m} + 2\ket{m+1} -\ket{m+2} \right),\\
\end{equation}
and the state $\ket{\overline{\psi}}$ to be measured is as follows:
\begin{align}
    \ket{\overline{\psi}} &= \dfrac{1}{3\sqrt{2+\gamma^2}} \left(   (\gamma+\sqrt{3}) (\ket{00}+\ket{11}+\ket{22}) + \right. \nonumber\\
    & \hspace{2.2cm} \gamma (\ket{01}+\ket{12}+\ket{20}) + \nonumber\\
    & \left. \hspace{2.2cm} (\gamma-\sqrt{3}) (\ket{02}+\ket{10}+\ket{21}) \right). \label{cglmpoptimal}
\end{align}
\noindent Here all additions performed inside the kets are modulo~3.

The above measurements and states shall then be our reference
system. To certify the state, as usual, Alice and Bob will each
attach a trusted qutrit to the entangled pair. The next step is to
construct CNOT operators with which to build a partial swap, see
Section \ref{method}. The key point is how to build the
translation operator $P$ from the measurement projectors defined
in Eq. (\ref{cglmpoptimal}). There are many choices to do so; we
chose the simplest combination:
\begin{align}\label{cglmpunitary}
P &= \sum_{i=0}^2 \overline{E}^0_i \left(-\frac{1}{2}\overline{E}^1_i - 2\overline{E}^1_{i+1} +\overline{E}^1_{i+2}\right)
\end{align}
which indeed is a translation operator mapping
$\ket{0}\rightarrow\ket{1}\rightarrow\ket{2}\rightarrow\ket{0}$. Here, addition of indices is modulo 3.
Since Alice and Bob's optimal operators are identical, the above
formula also applies to Bob's settings if we replace $E$'s by
$F$'s.

The choice above in Eq.~\eqref{cglmpunitary} is a valid unitary
operator only for the optimal strategy $E_a^0=\ketbra{a}{a}$ and
$E_a^1=\ketbra{\alpha_a}{\alpha_a}$. In a device independent
scenario there is no guarantee that this choice still defines a
valid unitary operator. We therefore introduce an extra auxiliary
(non-hermitian) operator, $\hat{P}_A$, with the constraint that $\hat{P}_A^\dagger
P(E^x_a)\geq 0$. For Bob's side, the swap operators are defined
exactly the same way as above for Alice. Thus, we require also
another auxiliary operator $\hat{P}_B$. We therefore use two
localizing matrices $\Gamma(\hat{P}_A^\dagger P(E^x_a),\cc)$,
$\Gamma(\hat{P}_B^\dagger P(F^x_a),\cc)$ to self-test the CGMLP
inequality. As mentioned in Section~\ref{generalSwap}, all three
semidefinite matrices can be taken real here, since, for any
feasible point $\overline{\cc}$ of the corresponding complex SDP,
the real vector $\Re\{\overline{\cc}\}$ returns the same state
fidelity, and the real matrices
$\Gamma(\Re\{\overline{\cc}\}),\bar{\Gamma}(\hat{P}_A^\dagger
P(E^x_a),\Re\{\overline{\cc}\}),\bar{\Gamma}(\hat{P}_B^\dagger
P(F^x_a),\Re\{\overline{\cc}\})$ are also positive semidefinite and
satisfy the appropriate linear constraints.

\begin{figure}[htbp!]
    \begin{center}
    \includegraphics[width=0.46\textwidth]{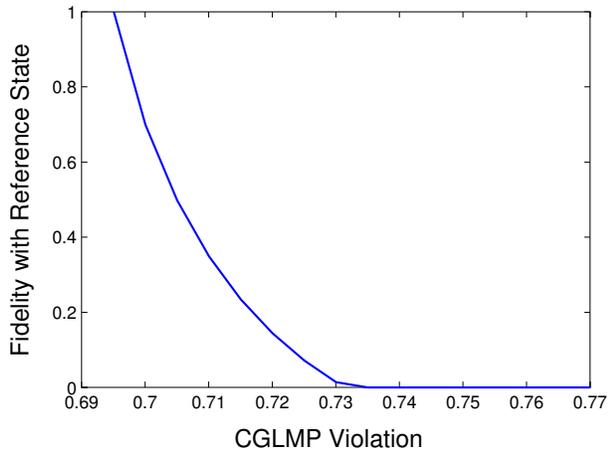}
    \caption{Minimum fidelity of the state swapped out the operators defined above. The blue line represents the minimum fidelity obtained from the SDP hierarchy. The hierarchy we used is the smallest hierarchy possible for the problem to be defined. \label{cglmpfidelity}}
\end{center}
\end{figure}

Figure \ref{cglmpfidelity} shows a plot of the minimum fidelity
with respect to the reference state $\ket{\overline{\psi}}$ as a
function of the CGLMP violation.

One can see that, as the violation tends to the quantum minimum
($\approx$ 0.6950), the fidelity of the general black box with
respect to the reference state (non maximally entangled state)
tends to 1. This shows that any quantum system violating the CGLMP
inequality maximally is equivalent via local isometries to the
non-maximally entangled state described in \cite{conject}. This
proves the conjecture that only non-maximally entangled states can
violate CGLMP maximally.

\subsection{Certification of entangled measurements}
\label{entangling}

In Ref.~\cite{rafael} an entanglement-swapping based protocol to
certify entangled measurements in a device-independent way is
presented. In this protocol, four parties, A, B, C$_A$, C$_B$ are
allowed to conduct one out of two possible dichotomic measurements
in their respective subsystems. By showing a maximal violation of
the CHSH inequality between parties A, C$_A$ and B, C$_B$, the state
shared by Alice and Bob is certified to be separable. Then parties
C$_A$, C$_B$ are brought together and a collective
measurement $C$ with four possible outcomes is conducted on the joint system C$_A$-C$_B$. As
shown in \cite{rafael}, if, conditioned on any of $C$'s outcome, Alice and Bob observe an absolute value for CHSH larger than $\sqrt{2}$, then measurement $C$ must be entangled.

Unfortunately, the previous protocol relies on the assumption that
parties A, C$_A$ or B, C$_B$ violate CHSH maximally. This assumption
is crucial to identify the degrees of freedom in parties C$_A$ and C$_B$
on which to test for the effect of measurement $C$. Indeed, the maximal violation
of both CHSHs identifies one qubit on each of these systems on which
the authors of~\cite{rafael} show that the action of measurement $C$ is entangled. Without
this identification, the experimental statistics can always be modeled with
a separable POVM between two parts C$_A$ and C$_B$ of the joint system C. Indeed, consider the situation where C$_A$ and C$_B$,
in addition to the degrees of freedom relevant for their CHSH measurement, share extra pairs of
maximally entangled states. Then, at the time of conducting measurement $C$,
C$_A$ can use the maximally
entangled states to teleport her internal state to C$_B$, who, in
turn, conducts measurement $C$ on his side. The resulting POVM
allows one to implement effectively measurement $C$ on C$_A$ and
C$_B$'s internal states, and is achievable via 1-way Local
Operations and Classical Communication (and is, hence, separable).
Therefore, we say that a measurement is entangled if one can identify
local degrees of freedom on which its action is entangled.

\begin{figure}[htbp!]
    \begin{center}
    \includegraphics[width=0.46\textwidth]{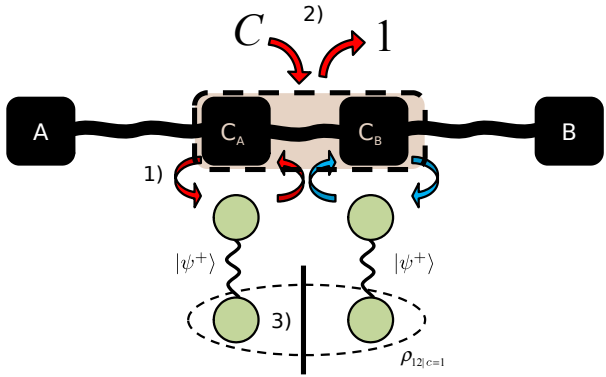}
    \caption{Certification of non-separability of the measurement $C$. The scheme works as follows: 1) We perform a (local) swap between the boxes C$_A$, C$_B$ and two particles, each maximally entangled with particles 1, 2, respectively. 2) We perform measurement $C$ and post-select on result $c=1$. 3) We minimize the negativity of the state $\rho_{12|c=1}$.}
    \label{entangle}
\end{center}
\end{figure}

In order to single out appropriate degrees of freedom where to test the
action of measurement $C$ for non-maximal CHSH violations, we make use of the SWAP tool (see Figure~\ref{entangle}).
Namely, we present a trusted qubit system to each of C$_A$ and C$_B$ that
we swap inside the two parties' boxes with the help of the unitary swaps $\SS_A$ and $\SS_B$. These
local unitaries define a qubit degree of freedom inside each party's system.
We then
test whether the action of measurement $C$ on these particular degrees of freedom
is entangled by analysing the state of two additional qubits (systems 1 and 2 in Figure~\ref{entangle}) initially choosen to be maximally entangled with the two swapped qubits.

We choose to analyse the state of these two qubits conditioned on one of the possible outcomes of the joint measurement $C$ (as opposed to conditioning on four possible outcomes as in~\cite{rafael}). It is thus sufficient to consider that $C$ can either produce the value $c=+1$ or $c=-1$, where $+1$ denotes `success' of the measurement, i.e. projection onto one Bell state, and $-1$ denotes projection onto one of the three remaining Bell states. The state of interest is then $\rho_{12|c=1}$.
Obviously, if $C$ can be described as a separable operator
on the degrees of freedom considered, $\rho_{12|c=1}$ cannot be entangled
at the end of the procedure. We can quantify the entanglement of the state
$\rho_{12|c=1}$ via its negativity $\mathcal{N}$
\cite{computable}, which, although non-linear in the components of
$\rho_{12|c=1}$, can be estimated via SDP~\cite{moroder}.

This negativity can be bounded as a function of the following three inequalities:
\begin{align}
\mathcal{B}_1 &= \langle A_0C_0^A\rangle + \langle A_0C_1^A\rangle + \langle A_1C_0^A\rangle - \langle A_1C_1^A\rangle \le 2,\nonumber\\
\mathcal{B}_2 &= \langle B_0C_0^B\rangle + \langle B_0C_1^B\rangle + \langle B_1C_0^B\rangle - \langle B_1C_1^B\rangle \le 2,\nonumber\\
\mathcal{B}_3 &= \langle (1+C)(A_0B_0 + A_0B_1 + A_1B_0 - A_1B_1)\rangle \nonumber\\
&\ \ \ -2\langle C \rangle \le 2,
\label{ABC}
\end{align}
by solving the following SDP program:
\begin{align}
    f^S=\min \;\; & \tr(\sigma_{-})\nonumber\\
    \textrm{s.t. }& \Gamma^S(\cc)\geq 0 \nonumber\\
                  & \rho_{12|c=1} = g(\cc)\\
                  & \rho_{12|c=1}^{T_1} =\sigma_{+}-\sigma_{-} \nonumber\\
                  & \sigma_{\pm}\geq 0 \nonumber\\
                  & \sum_i f_i^{B_j}(\cc_{i}) = \mathcal{B}_j\ ,\ j=1,2,3\ ,\nonumber
\end{align}
where $\rho_{12|c=1}$ is a function of the entries of the moment matrix $\Gamma^S(\cc)$ which we denote by $g(\cc)$. The optimization runs over the moments $\cc$ and the hermitian matrices $\sigma_{\pm}$. Above, we made use of the variational formula for the negativity of a bipartite state~\cite{computable}. Note that $T_{1}$ denotes the partial transpose with respect to the first subsystem (c.f. also~\cite{moroder}), and $f_i^{B_j}$ are the coefficients of the Bell inequality $\mathcal{B}_j$.

\begin{figure}[htbp!]
    \begin{center}
    \includegraphics[scale=0.66]{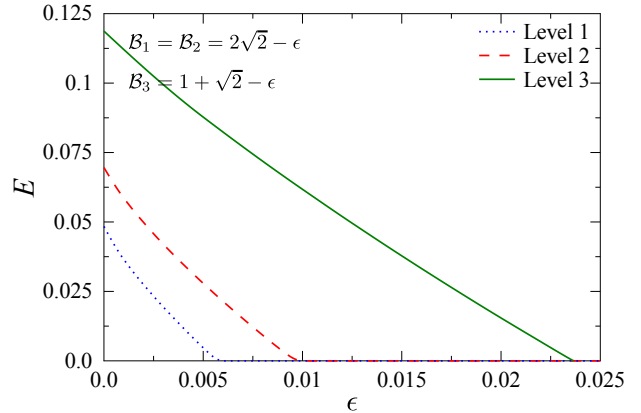}
    \caption{Certification of non-separability of measurement $C$. We minimize the quantity
    $E =P(c=1)\mathcal{N}\left(\rho_{12|c=1}\right)$ as a function of the value $\epsilon$,
    where $\mathcal{N}\left(\rho_{12|c=1}\right)$ denotes the negativity of the state $\rho_{12|c=1}$.
    The curves correspond to different levels of the NPA hierarchy \cite{navascues,navascueslong}.
    $E>0$ signals that Charlie's measurement $C$ is entangled.}
    \label{measneg}
\end{center}
\end{figure}

Figure~\ref{measneg} shows the minimum value of $E\equiv
P(c=1)\mathcal{N}\left(\rho_{12|c=1}\right)$ when the Bell violations are chosen to be the ones expected from perfect measurements on two Werner states $\rho=V\ket{\psi^-}\bra{\psi^-}+(1-V)\openone/4$ with $\epsilon=2\sqrt{2}(1-V)$ up to first order in $\epsilon$, i.e. $\mathcal{B}_1 = \mathcal{B}_2 = 2\sqrt 2 -\epsilon$, $\mathcal{B}_3 = 1+\sqrt 2 -\epsilon.$

The values in Figure~\ref{measneg} were obtained by using the SeDuMi \cite{sedumi} package in Matlab. The three distinct curves
correspond to different relaxations of the SDP problem with
respective dimensions 144, 168, and 200 of the moment matrix $\Gamma^S(c)$. At
the highest NPA level 3 (corresponding to a 200 dimensional moment
matrix), we find that for
$\epsilon<0.023$ the action of measurement $C$ on the degrees
of freedom identified by the SWAPs cannot be described by a
separable POVM. In the figure, we may also observe that as
$\epsilon$ goes to zero measurement $C$
tends to act as a Bell state measurement providing (close to)
$E=1/8$.


\subsection{SWAP and maximally mixed states: work extraction and dimension estimates}
\label{work}

The relation between work and information has been a matter of
scientific debate since the dawn of statistical mechanics. In this
regard, it has been shown in \cite{szilard, oscar} that the
knowledge of the state of a quantum system, as measured by the
smooth min and max entropies, can be used to generate work. Since
the SWAP tool allows us to acquire knowledge about the state
inside the box, it is hence not surprising that it can also lower
bound its potential for work extraction.

\begin{figure}[htbp!]
    \begin{center}
    \includegraphics[scale=0.4]{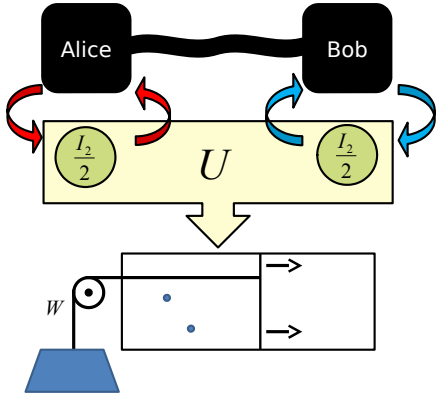}
    \caption{Non-locality and work. We let the boxes interact unitarily with two maximally mixed qubits, representing the position of two particles in a Szilard engine. The resulting pure state can then be rotated so that both particles end up on the left side of the engine with very high probability, hence producing storable work.}
    \label{work_CHSH}
\end{center}
\end{figure}

\begin{figure}[htbp!]
    \begin{center}
    \includegraphics[scale=0.93]{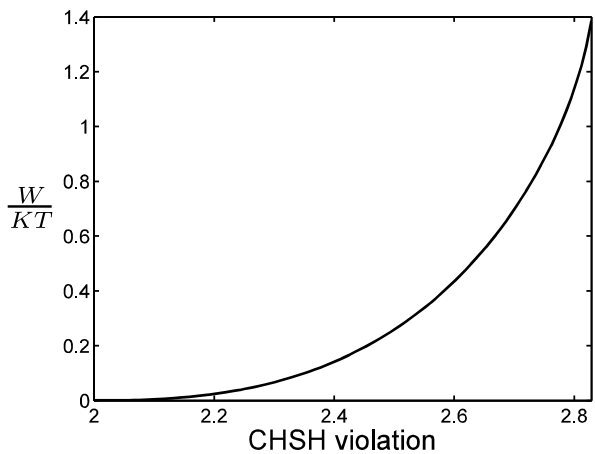}
    \caption{Extractable work per $KT$ as a function of the CHSH violation of an isotropic box.}
    \label{work_CHSH2}
\end{center}
\end{figure}

In Figure \ref{work_CHSH} we use local swaps over two maximally
mixed trusted qubits initially inside a Szilard engine in contact
with a bath at temperature $T$. Since by definition the state of a
maximally mixed qubit is completely unknown, such states can be
considered as a free resource. After the interaction with the box,
the state of the two qubits gets purified, and they can be rotated
to one of the sides of the box with high probability, thus pulling
a weight. Under the assumption that the energy operator of the
system inside the box is fully degenerate, it can be proven that
the amount of work extracted in this way is related to the
difference between the maximum and minimum eigenvalues of
$\rho_{\mbox{swap}}$, see Appendix \ref{work_app}. This difference
can be estimated via semidefinite programming. Figure
\ref{work_CHSH2} shows a plot of the CHSH violation vs the
minimum amount of work extractable via this scheme. Notice that
work extraction is possible as long as the system is non-local.
Also, for the maximal CHSH violation, the amount of work
extractable per $KT$ is $W/KT=2\ln(2)$, the work content of two
pure qubits \cite{work_pure, work_pure2}.

In more complex non-locality scenarios, similar optimizations over the spectrum of the swapped state could also give clues about the dimensionality of the
state inside Alice and Bob's boxes. In
\cite{noisy_op}, it is proven that a state $\rho\in B(\H)$ can be
transformed into a state $\sigma\in B(\H)$ by means of unitary
transformations over $\rho$ and maximally mixed ancillas iff
$\rho$ majorizes $\sigma$, i.e., if
$\sum_{i=1}^k\lambda_i(\rho)\geq \sum_{i=1}^k\lambda_i(\sigma)$,
for all $k$ (here $\lambda_i(M)$ denotes the $i^{th}$ greatest
eigenvalue of the matrix $M$). Now, let $\rho_{\mbox{box}}\in
B(\C^D)$ be the state inside the box(es), and consider the
transformation $\rho=\rho_{\mbox{box}}\otimes
(\id_d/d)\to\sigma=(\id/D)\otimes\rho_{\mbox{swap}}$ that results
from applying a swap operator between the box and a trusted qudit
in the maximally mixed state, followed by the replacement of the
state inside the box by a maximally mixed qudit. From
\cite{noisy_op} it follows that
$\lambda_{1}(\rho)/d\geq\lambda_{1}(\rho_{\mbox{swap}})/D$, and so
$D$ satisfies \be
D\geq d\lambda_{1}(\rho_\text{swap}).
\ee

\noindent By minimizing $\lambda_1(\rho_\text{swap})$, we could
hence (in principle) lower bound the dimensionality $D$ of the system inside the
box.

\section{Conclusion}

To appreciate and ultimately certify a quantum experiment, a
considerable amount of detailed information about the internal workings of the experimental setup 
is generally needed. For
instance, in typical ion-trap experiments \cite{ion} the state of
the ions is considered to be (approximately) embedded in a
two-dimensional Hilbert space, and the behavior of the ion when
subject to sequential laser pulses is well known. Without this
knowledge, it would be difficult to make sense of most ion-trap
experiments. Remarkably, there exist situations where this kind of 
information is not required in order to take advantage of a
(possibly complex) system.

In this paper we have introduced the SWAP tool to characterize
quantum systems in arbitrary Bell-type scenarios. We showed that
it provides much stronger bounds for known self-testing
procedures, compared to previously-known techniques. We also
showed that it provides robust self-testing in new contexts, such
as the self-testing of qutrit states and of entangled
measurements. Finally, we used the SWAP tool to relate nonlocal
correlations to work extraction and quantum dimension,
demonstrating that the tool can be used to study a wide variety of
properties. All of these results are naturally noise tolerant and
were obtained from the sole knowledge of accessible statistics.

The generality of the method exposed here raises new questions:
How far can one go with the device-independent approach? Are
Hilbert spaces necessary at all? Could the whole field of quantum
information science be reformulated as a theory of black boxes?

\section*{Acknowledgements}
This work is funded by the Singapore Ministry of Education (partly
through the Academic Research Fund Tier 3 MOE2012-T3-1-009) and by
the National Research Foundation of Singapore. M.N. acknowledges
support from the John Templeton Foundation, the ERC Advanced Grant
NLST,  EPSRC DIQIP, the European Commission (EC) STREP ``RAQUEL''
and the MINECO project FIS2008-01236, with the support of FEDER
funds. T.V. acknowledges financial support from a J\'anos Bolyai
Grant of the Hungarian Academy of Sciences, the Hungarian National
Research Fund OTKA (PD101461), and the
T\'AMOP-4.2.2.C-11/1/KONV-2012-0001 project.

\begin{appendix}

\section{Proof of relation (\ref{rela})}
\label{relation}

In order to relate both quantities, we express $\ket{\psi}$ in the Schmidt decomposition

\be
\ket{\psi}=\sum_{i\geq 1}\sqrt{p_i}\ket{\psi_i}_{A'B'}\ket{\phi_i}_{AB}.
\ee

A singlet fidelity $F$ with respect to $A'B'$ hence implies:

\be
\sum_{i\geq 1} p_i|\braket{\overline{\psi}}{\psi_i}|^2=F.
\ee

\noindent Notice that the left hand side is upperbounded by $p_1$ (assuming that the Schmidt coefficients are ordered decreasingly, i.e., that $p_1$ is the maximum).

On the other hand, we have that

\be
p_1|\braket{\overline{\psi}}{\psi_1}|^2+1-p_1\geq \sum_i p_i|\braket{\overline{\psi}}{\psi_i}|^2=F.
\ee

\noindent Combining the last inequality with $p_1\geq F$, we get that

\be
p_1|\braket{\overline{\psi}}{\psi_1}|^2\geq 2F-1.
\ee

Finally, note that

\begin{eqnarray}
s^2&&=\max_{\ket{junk}}|\bra{\psi}(\ket{\overline{\psi}}\otimes\ket{junk})|^2\nonumber\\
&&\geq p_1|\braket{\psi_1}{\overline{\psi}}|^2\geq 2F-1,
\end{eqnarray}

\noindent where the inequality follows from taking $\ket{\mbox{junk}}=\ket{\phi_1}$.

\section{Non-Maximally Entangled State Certification with the tilted CHSH inequality}
\label{tilted_app}

For convenience, we shall work in the local basis such that Bob's optimal measurements are $\overline{B}_0=\sigma_z$ and $\overline{B}_1=\cos(2\mu)\sigma_z-\sin(2\mu)\sigma_x$. Note that $\bar{B}_1$ is no longer $\sigma_x$, as in the CHSH case, thus the swap operator defined in (\ref{uvu}) no longer works for Bob. We will have to construct the analog of $\sigma_x$ by combining the operators $B_0$ and $B_1$. We thus introduce an auxiliary dichotomic operator $B_2$, with $B_2^2=\id$, and impose relations between $B_0,B_1,B_2$ such that $B_2$ behaves as if it is $(\cos(2\mu)B_0-B_1)/\sin(2\mu)$. Following Section \ref{method}, the appropriate constraint is
\begin{align}
    B_2 \dfrac{\cos(2\mu)B_0-B_1}{\sin(2\mu)} \geq 0. \label{auxiliaryconstraint}
\end{align}
This will force $B_2$ to share the same eigenvectors as $(\cos(2\mu)B_0-B_1)/\sin(2\mu)$, and will identify both operators in the optimal case.

The swap operators in this case will then be $\SS_B=U'_BV'_BU_B'$, with
\begin{align}
    &U'_B = \left(\mathbb{I}\otimes\ketbra{0}{0}+B_2\otimes\ketbra{1}{1} \right),\nonumber\\
    &V_B'=\left(  \dfrac{\mathbb{I}+B_0}{2}\otimes \mathbb{I} + \dfrac{\mathbb{I}-B_0}{2}\otimes \sigma_x \right), \label{swapoperatorstilted}
\end{align}
while $\SS_A$ is the same as in Eq. (\ref{uvu}).

With the introduction of this extra auxiliary operator, our moment matrix is now enlarged with effectively two operators on Alice's side $(A_0,A_1)$ and three operators on Bob's side $(B_0,B_1,B_2)$. The constraint (\ref{auxiliaryconstraint}) is enforced by imposing that the localizing matrix defined by
\begin{equation}
\begin{split}
\Gamma&\left(B_2\dfrac{\cos(2\mu)B_0-B_1}{\sin(2\mu)}\right)_{ss'}\\
&=\frac{\cos(2\mu)}{\sin(2\mu)}\cc_{s^\dagger B_2 B_0 s'}-\frac{1}{\sin(2\mu)}\cc_{s^\dagger B_2 B_1 s'}
\end{split}
\end{equation}
is positive semidefinite.

\section{Work extraction}
\label{work_app}

Let $\rho_{AB}$, with spectral decomposition $\rho_{AB}=\sum_{k=1}^4\lambda_k\proj{\psi_k}$, describe the state of two particles in a Szilard engine of length $L$ and area $A$, with $\lambda_i\geq\lambda_{i+1}$. Denoting by $L$ ($R$) the state corresponding to a particle being on the left (right) of the Szilard engine, we can always apply a unitary $U$ over the state $\rho_{AB}$ such that $U\ket{\psi_1}=\ket{L,L}$, $U\ket{\psi_2}=\ket{R,L}$, $U\ket{\psi_3}=\ket{L,R}$, $U\ket{\psi_4}=\ket{R,R}$. It follows that a population of $N$ particle pairs in a Szilard engine, initially in state $\rho_{AB}^{\otimes N}$, can be brought to a situation where $N_L=N(\lambda_1-\lambda_4+1)$ particles are on the left side of the engine, and $N_R=N(\lambda_4-\lambda_1+1)$ are on the right side. If we place a movable wall connected to a weight on the left side of the engine (in contact with a bath at temperature $T$), the pressure over the wall is equal to

\be
P=\frac{N_LKT}{Az}-\frac{N_RKT}{A(L-z)},
\ee

\noindent where $z$ denotes the position of the piston. At constant temperature, the equilibrium position of the piston is $z_{\mbox{eq}}=\frac{N_L L}{N_L+N_R}$. The work extracted in the process of moving the piston from $z=L/2$ to $z_{\mbox{eq}}$ is

\begin{eqnarray}
&W=\int_{L/2}^{z_{\mbox{eq}}}PAdz=\nonumber\\
&=KT\left\{N_L\ln\left(\frac{2N_L}{N_L+N_R}\right)+N_R\ln\left(\frac{2N_R}{N_L+N_R}\right)\right\}=:\nonumber\\
&:=W(\lambda_1-\lambda_4).
\end{eqnarray}

Finally, the minimum value of $\lambda_4-\lambda_1$ can be extracted from $\rho_{\mbox{swap}}$ via the following SDP:

\begin{equation}
\begin{split}
\min\ &\mu_1-\mu_4\nonumber\\
\mbox{s.t.}\ &\rho_{\mbox{swap}}-\mu_4\id\geq 0,\nonumber\\
& \mu_1\id-\rho_{\mbox{swap}}\geq 0.
\end{split}
\end{equation}

Since
\begin{eqnarray}
&\rho_{\mbox{swap}}-\mu_4\id \geq 0\Rightarrow\Re\{\rho_{\mbox{swap}}\}-\mu_4\id\geq 0,\nonumber\\
&\mu_1\id-\rho_{\mbox{swap}} \geq 0\Rightarrow \mu_1\id-\Re\{\rho_{\mbox{swap}}\}\geq 0,
\end{eqnarray}
it follows that the free variables in the corresponding SDP can be taken real, as in the previous examples.

\end{appendix}


\begin{thebibliography}{99}
\bibitem{endnotedet} Here and throughout the whole text, we consider only asymptotic statements, i.e. we assume that the measurement statistics are perfectly reconstructed. Also, it is understood that the observed Bell violation is not fake: the detection loophole must be closed and no-signaling must be assumed or guaranteed.
\bibitem{endnoteuv} Having chosen the ancillas $A'$ and $B'$ in the $\ket{0}$ state, we could have used ${\cal S}_{AA'}=U_{AA'} V_{AA'}$ and the same for Bob, because the first $U$ operator acts trivially. This is effectively equivalent to the isometry used in~\cite{mckague}: therefore, the enormous difference in robustness between that result and ours shows the power of the SDP tool.
\end{thebibliography}
\end{document}